\documentclass[11pt]{article}
\usepackage{geometry}                
\usepackage[parfill]{parskip}    
\usepackage{graphicx}
\usepackage{amssymb,amsmath,amsthm}
\usepackage{epstopdf,color}
\usepackage[shortlabels]{enumitem}
\author{David Degras$^{\rm a}$ and Martin A. Lindquist$^{\rm b}$ 
\\\\
$^{\rm a}${\it{Department of Mathematical Sciences, DePaul University, USA.}}\\ 
$^{\rm b}${\it{Department of Biostatistics, Johns Hopkins University, USA.}}\\ 
}

\begin{document}
\title{A Hierarchical Model for Simultaneous Detection and Estimation in Multi-subject fMRI Studies}
\author{David Degras$^{\rm a}$ and Martin A. Lindquist$^{\rm b} \footnote{Corresponding author: 615 N. Wolfe Street, E3634; Baltimore, MD 21205, e-mail: mlindqui@jhsph.edu.  This research was supported by NIH grant R01EB016061.}$ 
\\\\
$^{\rm a}${\it{Department of Mathematical Sciences, DePaul University, USA.}}\\ 
$^{\rm b}${\it{Department of Biostatistics, Johns Hopkins University, USA.}}\\ 
}

\maketitle 
\thispagestyle{empty}
\newpage

\thispagestyle{empty}

\newcommand{\balpha}{\boldsymbol{\alpha}}
\newcommand{\bbeta}{\boldsymbol{\beta}}
\newcommand{\bdelta}{\boldsymbol{\delta}}
\newcommand{\bgamma}{\boldsymbol{\gamma}}
\newcommand{\bGamma}{\boldsymbol{\Gamma}}
\newcommand{\bGammainv}{\boldsymbol{\Gamma}^{-1}}
\newcommand{\boldeta}{\boldsymbol{\eta}}
\newcommand{\bphi}{\boldsymbol{\varphi}}
\newcommand{\bA}{\mathbf{A}}
\newcommand{\bb}{\mathbf{b}}
\newcommand{\bc}{\mathbf{c}}
\newcommand{\bd}{\mathbf{d}}
\newcommand{\rmd}{\mathrm{d}}
\newcommand{\bD}{\mathbf{D}}
\newcommand{\bdj}{\mathbf{d}_j}
\newcommand{\bg}{\mathbf{g}}
\newcommand{\bh}{\mathbf{h}}
\newcommand{\bH}{\mathbf{H}}
\newcommand{\bI}{\mathbf{I}}
\newcommand{\bS}{\mathbf{S}}
\newcommand{\bT}{\mathbf{T}}
\newcommand{\bM}{\mathbf{M}}
\newcommand{\bMj}{\mathbf{M}_j}
\newcommand{\bN}{\mathbf{N}}
\newcommand{\bNj}{\mathbf{N}_j}
\newcommand{\bP}{\mathbf{P}}
\newcommand{\bRj}{\mathbf{R}_j}
\newcommand{\btildeRj}{\tilde{\mathbf{R}}_j}
\newcommand{\brj}{\mathbf{r}_j}
\newcommand{\bU}{\mathbf{U}}
\newcommand{\bV}{\mathbf{V}}
\newcommand{\bVj}{\mathbf{V}_j}
\newcommand{\hatVjinv}{\hat{\mathbf{V}}_j^{-1}}
\newcommand{\bW}{\mathbf{W}}
\newcommand{\bXj}{\mathbf{X}_j}
\newcommand{\bXjl}{\mathbf{X}_{jl}}
\newcommand{\by}{\mathbf{y}}
\newcommand{\byj}{\mathbf{y}_j}
\newcommand{\bzero}{\mathbf{0}}
\newcommand{\bPhij}{\boldsymbol{\Phi}_j}
\newcommand{\bxij}{\boldsymbol{\xi}_j}
\newcommand{\bxi}{\boldsymbol{\xi}}
\newcommand{\bxijl}{\boldsymbol{\xi}_{jl}}
\newcommand{\bepsilonj}{\boldsymbol{\varepsilon}_j}
\newcommand{\bepsilon}{\boldsymbol{\varepsilon}}
\newcommand{\hatepsilonj}{\hat{\boldsymbol{\varepsilon}}_j}
\newcommand{\tr}{\mathrm{tr}}
\newcommand{\cov}{\mathrm{Cov}}
\newcommand{\var}{\mathrm{Var}}

\begin{abstract}
In this paper we introduce a new hierarchical model for the simultaneous detection of brain activation and estimation of the shape of the hemodynamic response in multi-subject fMRI studies. The proposed approach circumvents a major stumbling block in standard multi-subject fMRI data analysis, in that it both allows the shape of the hemodynamic response function to vary across region and subjects, while still providing a straightforward way to estimate population-level activation. An efficient estimation algorithm is presented, as is an inferential framework that not only allows for tests of activation, but also for tests for deviations from some canonical shape. The model is validated through simulations and application to a multi-subject fMRI study of thermal pain. 

\end{abstract}
\newpage

\section{INTRODUCTION}

Depending on their scientific goals, researchers in functional magnetic resonance imaging (fMRI) often choose modeling strategies with the intent to either {\em detect} the magnitude of activation in a certain brain region, or {\em estimate} the shape of the hemodynamic response associated with the task being performed \cite{poldrack2011handbook}.  While most of the focus in neuroimaging to date has been on detection \cite{lindquist2008statistical}, the magnitude of evoked activation cannot be accurately measured without either assuming or measuring timing and shape information as well. In practice, many statistical models of fMRI data attempt to simultaneously incorporate information about the shape, timing, and magnitude of task-evoked hemodynamic responses.

As an example, consider the general linear model (GLM) approach \cite{worsley}, which is arguably the dominant approach towards analyzing fMRI data. It models the fMRI time series as a linear combination of several different signal components and tests whether activity in a brain region is related to any of them. Typically the shape of the hemodynamic response is assumed {\it a priori}, using a canonical hemodynamic response function (HRF) \cite{friston1998event, glover1999deconvolution}, and the focus of the analysis is on obtaining the magnitude of the response across the brain. However, it is well-known that the shape of the HRF varies both across space and subjects  \cite{aguirre, Schacter,Handwerker2004,Badillo2013}; thus assuming a constant shape across all voxels and subjects may give rise to significant bias in large parts of the brain \cite{lindquist6, lindquist8}. The constant HRF assumption can be relaxed by expressing it as a linear combination of several known basis functions. This can be done within the GLM framework by convolving the same stimulus function with multiple canonical waveforms and including them as multiple columns of the design matrix for each condition. The coefficients for an event type constructed using different basis functions can then be combined to fit the evoked HRF in that particular area of the brain. 

The ability to use basis sets to capture variations in hemodynamic responses depends both on the number and shape of the reference waveforms that are used in the model. For example, the finite impulse response (FIR) basis set, consists of one free parameter for every time-point following stimulation in every cognitive event-type that is modeled \cite{glover1999deconvolution, goutte}. Therefore, it can be used to estimate HRFs of arbitrary shape for each event type in every voxel of the brain. Another, perhaps more common approach, is to use the canonical HRF together with its temporal and dispersion derivatives to allow for small shifts in both the onset and width of the HRF. Other choices of basis sets include those composed of principal components \cite{aguirre, woolrich}, cosine functions \cite{zarahn2002using}, radial basis functions \cite{riera}, spectral basis sets \cite{Liao02} and inverse logit functions \cite{lindquist6}. For a critical evaluation of a number of commonly used basis sets, see \cite{lindquist6} and \cite{lindquist8}.

Though basis sets allow the constant HRF assumption to be relaxed in the GLM framework, they are still not without problems, particularly when performing multi-subject analysis. Most analyses of multi-subject fMRI data involve two separate models. A first-level GLM is performed separately on each subject's data, providing subject-specific contrasts of parameter estimates. A second-level model is thereafter used to provide population inference on whether the contrasts are significantly different from zero and assess the effects of any group-level predictors, such as group status or behavioral performance. However, it is problematic to define appropriate first-level contrasts that truly capture the behavior we are interested in detecting, when multiple basis sets are included in the model. For example, when each condition consists of multiple basis functions it is not self-evident how to properly define a relevant contrast comparing the difference in activation between the two conditions.

There have been a number of suggestions about how to deal with this issue, most notably using only the ``main'' basis set \cite{kiehl2001event}. In the case of the canonical HRF and its derivatives, this entails only using the coefficient corresponding to the canonical HRF, and treating the coefficients corresponding to the derivatives as nuisance parameters. While this works relatively well for small deviations from the HRFs canonical form, it quickly falls apart as the shape begins to differ. To circumvent this problem \cite{Calhoun04} suggested using the norm of the coefficients for the canonical HRF and its derivatives. Another suggestion \cite{lindquist8} is to re-create the HRF after estimation and use the resulting amplitude as the contrast of interest.

In this work we introduce a new approach towards multi-subject analysis of fMRI data that enables 
us to simultaneously estimate the specific shape of the HRF for a subject in a given voxel, and to obtain a population-level estimate of the magnitude of activation. 
 This approach offers the flexibility of basis sets  while retaining the simplicity of multi-subject inference with a single canonical HRF. 
 We also provide an inferential framework that allows us to test both for activations as well as for any differences in HRF shape from some canonical form.

The idea of performing joint estimation and detection is not new in the neuroimaging literature. For example, Makni and colleagues \cite{makni2005joint,makni2008fully} have suggested a Bayesian approach towards the detection of brain activity that uses a mixture of two Gaussian distributions as a prior on a latent neural response, whereas the hemodynamic impulse response is constrained to be smooth using a Gaussian prior. In this model all parameters of interest are estimated from the posterior distribution using Gibbs sampling. Later work has provided a number of interesting extensions of this model, including to spatial mixture models \cite{vincent2010spatially} and reformulating it as a missing data problem that allows for a simplified estimation procedure \cite{chaari2012fast}.

Our suggested model takes a different approach. Here the HRF is modeled as a linear combination of B-spline functions 
(see e.g., \cite{Genovese2000} for an early use of spline functions in model fitting). 
We assume that subject-level HRFs are random draws from a population-level distribution and that for any given voxel the population average response across stimuli will vary only in scale. We provide an efficient algorithm for estimating the model parameters as well as inferential methods. 
The latter includes both tests of activation and for deviations in the HRF from some canonical form.

This paper is organized as follows. In Section 2 we introduce our hierarchical model. In Sections 3 and 4 we outline an efficient algorithm for estimating the model parameters and performing inference on them. In Sections 5 and 6 we evaluate the performance of the model on a series of simulated data sets and data from an experiment studying the effects of thermal pain.  Both the simulated and real data were previously used in a large study of flexible HRF modeling procedures \cite{lindquist8}. The proposed method is shown to outperform each of the previously tested approaches, which include the canonical GLM plus its derivatives, the smooth FIR model, and the inverse logit model.  We conclude with a discussion of the suggested approach.

\section{METHODS}

\subsection{The Hierarchical Model}

In this section we outline the proposed hierarchical model for simultaneous estimation and detection. 
For simplicity, we assume only one session per subject and a single  experimental group. 
To study multiple groups, it suffices to  apply the  model below and the estimation procedure of section \ref{sec: estimation} separately to each group.   
Similarly, the inference method of section \ref{sec: inference} easily extends to multiple samples. 
We assume that all scans have been acquired at the same repetition time $\Delta$ 
and registered to a  standard stereotactic space. For the $j^{th}$ subject ($1\le j \le n$), we model the BOLD response 
at the  $v^{th}$ voxel ($1 \le v \le V$) and $t^{th}$ scan ($1\le t \le T_j$) as  
 \begin{equation}\label{model}
\underbrace{y_{j}(v,t)}_{\textrm{BOLD}}  =\sum_{l=1}^L \int_0^{t\Delta}  \underbrace{s_{jl} (t\Delta -\tau)}_{\textrm{stimulus}}\,  \underbrace{  h_{j l }(v,\tau) }_{\textrm{HRF}} d\tau+
 \sum_{\nu=1}^q d_{j\nu}(v)  \underbrace{  \varphi_{j\nu}(t) }_{\textrm{nuisance} }  + 
  \underbrace{\varepsilon_{j}(v,t)}_{\textrm{noise}}   .
\end{equation}
Here the response consists of a linear combination of stimulus-induced signal of interest (represented by the first sum on the right-hand side of the equation), nuisance signal (the second sum) and noise. 
The stimulus function $s_{j l }$ is a stick-function that has baseline at zero and takes the value one whenever stimuli of the $ l^{th} $  type are presented. 
The nuisance signals $\varphi_{j\nu}, \, 1\le \nu \le q,$ typically include scanner drift, represented by polynomial and/or cosine basis sets, and physiological noise such as  head motion, heart beat, and respiration. The corresponding nuisance parameters $d_{j\nu}$ must  be estimated from the data.
The subject-level HRF  $h_{j l }$  decomposes into 
\begin{subequations}
\begin{equation}\label{hrf model decomp} 
h_{j l }(v, \tau) = h_{ l }(v, \tau) + \eta_{j l }(v,\tau) ,
\end{equation}
where $h_l$ is a population-level HRF (fixed effect) and $\eta_{j l }$ is a  subject-specific (random) effect. 
Hence, each subject-level HRF is assumed to be a random draw from a population with mean $h_{l}$.
Thus, similar to other recent methods (e.g. \cite{sanyal2012bayesian} and \cite{Zhang2013}) this allows us to borrow strength across subjects to improve subject-specific estimation.
The functions  $h_l(v, \cdot)$ and $\eta_{j l }(v, \cdot)$ are represented using a set of basis functions $(B_k )_{1\le k \le K}$: 
\begin{equation}\label{hrf model function basis}
  h_{ l }(v,\tau) = \sum_{k=1}^K  \gamma_{  l  k} (v) B_{k}(\tau), \quad
 \eta_{j l }(v,\tau) = \sum_{k=1}^K  \xi_{j l  k} (v) B_{k}(\tau),
\end{equation}
so estimation reduces to solving for the coefficients $\gamma_{  l  k} (v)$ and $\xi_{j l  k} (v)$. 
In practice we suggest specifying the $B_k$ as B-splines with regularly spaced knots over a time interval where the HRF is believed to be non-zero, say in the range between $0$ and $30$ seconds.   
B-spline basis sets have several desirable features. First, the coefficients of a function in a B-spline basis are very close to the function itself, i.e., the function values at the knots (possibly up to a scaling factor).
For this reason, B-spline coefficients are immediately interpretable and  
inference of local features of the HRF is greatly facilitated. In addition, the compact support
of B-splines typically induces sparsity in the design matrices and thus reduces the computational load.

Noting that the shape of the HRF at a  given brain location is mostly  determined  by physiological factors 
that are independent of the nature of the stimulus, we further assume that the population-level HRFs $h_l(v,\cdot), \, 1\le l \le L,$ 
have the same shape for all conditions and differ only in scale: 
\begin{equation}\label{simplification hrf}
\gamma_{  l  k}(v) = \beta_{ l } (v)\, \gamma_{k}(v) .
\end{equation}
Here the terms $\gamma_k (v), \, 1\le k \le K,$ determine the shape of the HRF, while $\beta_l (v)$ determines the amplitude of the response to stimulus of the $l^{th}$ type.
 To make these parameters identifiable, we impose the scale constraint  $\sum_k \gamma_{k}(v)^2=1$ 
and the orientation constraint $\gamma_{k_0}(v)>0$, with $k_0 = \mathrm{arg\,max}_k |\gamma_{k}(v)|$.  
Assumption \eqref{simplification hrf} considerably reduces the number of HRF parameters from $KL$ to $K+L$ 
(the same modeling assumption is used in e.g., \cite{makni2005joint}), 
 which allows one to use a reasonably large number of basis  functions while maintaining a good estimation accuracy. 
On the other hand,  the product form of  \eqref{simplification hrf} 
 makes model \eqref{model} nonlinear with respect to the parameters $\beta_l(v)$ and $\gamma_k(v)$. 
Note that the shape coefficients $\gamma_{k}(v)$ are well defined only if at least one of the HRFs $h_{l}(v,\cdot)$ 
is not identically zero, that is, if at least one experimental condition induces an activation at voxel $v$. 

\end{subequations}

The random effects $ \xi_{j  l  k} (v)$ in  \eqref{hrf model function basis}  represent subject-specific deviations from the group-level HRF coefficients $ \gamma_{ l  k} (v)$. They are assumed to be independent across  subjects and conditions.
In addition they are assumed to be Gaussian random vectors with mean zero and 
correlation structure that is stationary in time and constant in space: 
\begin{equation}\label{cov paradigm-related random effects}
\begin{array}{l}
 \mathrm{Cov}( \xi_{j   l  k}(v), \xi_{j'  l ' k'}(v)) =  (  \delta_{jj'} \delta_{ l   l '})   \,  \sigma_{\xi l}^2(v) \, \rho_{\xi  l }(|k-k'|) . 
 \end{array}
\end{equation}
In the previous equation, 
$\sigma_{\xi l}^2(v)$ denotes the common variance of the $ \xi_{j  l  k} (v)$  ($1\le j \le n,\, 1\le k \le K$), 
$\rho_{\xi l}$ is an autocorrelation function, and $\delta$ is  the Kronecker delta 
($\delta_{xy}=1$ if $x=y$, $\delta_{xy}=0$ if $x\ne y$). 
Note that $\rho_{\xi l }(0) = 1$.

To specify the  dependence structure of the noise component $\varepsilon$, 
we consider a partition of the spatial domain $\mathcal{D}$ into neuro-anatomic parcels $\mathcal{D}_1,\ldots,\mathcal{D}_M$ 
(e.g., Brodmann areas or any suitable brain atlas). 
We assume that for each voxel $v$ of a parcel $\mathcal{D}_m  $, 
 $\varepsilon_{j}(v,\cdot)$ is a stationary Gaussian AR($p$) process whose variance $\sigma_{\varepsilon m}^2$ and 
structural parameters $\boldsymbol{\theta}_{\varepsilon m}=(\theta_{\varepsilon 1m} ,\ldots, \theta_{\varepsilon pm})$ 
 only depend on $m$. In other words, 
  $\varepsilon_{j}(v,t) = \sum_{k=1}^p \theta_{km} \,\varepsilon_j(v,t-k) + e_j(v,t)$, 
  where the $e_j(v,t), \, 1\le t \le N_j ,$ are i.i.d. normal innovations. 
 The specification of the noise dependence at the parcel level 
reflects the belief that this noise is spatially smooth. 
An alternative way to characterize the spatial smoothness of $\varepsilon$ 
would be to model the AR parameters $\sigma_{\varepsilon }^2$  and $\boldsymbol{\theta}_{\varepsilon}$ 
as smooth functions of $v$.

\subsection*{Model Summary}

For the $j^{th}$ subject and $v^{th}$ voxel, the BOLD time course 
$\mathbf{y}_{j} (v) = \left( y_{j}(v,1),\ldots, y_{j}(v,T_j)\right)' $ can be expressed in matrix form as 
\begin{equation}\label{time course model}
\byj(v) = \bXj (\bbeta (v) \otimes \bgamma(v)) +\bXj \bxij (v) + \bPhij \bdj (v)+ \bepsilonj(v) . 
\end{equation}
As in the standard GLM, the design matrix $\bXj$ is the convolution of the stimulus functions $s_{jl}$ with the basis functions $B_k$. 
To be precise, $\bXj = ( \mathbf{X}_{j1},\ldots,\mathbf{X}_{jL} ) $ with  $\bXjl = 
 ( \mathbf{x}_{j1l },\ldots,\mathbf{x}_{jKl} ) $ and 
$\mathbf{x}_{jkl } = \big( \int_0^{\Delta} B_k(\tau) s_{jl}(\Delta-\tau)d\tau,\ldots, \int_0^{T_j  \Delta} B_k(\tau) s_{jl}(T_j \Delta -\tau)d\tau\big)'$. 
We have also written  
$\bbeta(v) = ( \beta_{1}(v), \ldots , \beta_{L}(v) )' $ and 
$\bgamma(v) = ( \gamma_{1}(v), \ldots , \gamma_{K}(v) )'$ for the amplitude and scale  parameters of the population-level HRF;  
$ \bxij (v) =  ( \bxi_{j1}(v)',\ldots,\bxi_{jL}(v)' )'$ and $\bxijl(v) =  ( \xi_{j1l}(v),\ldots,\xi_{jKl}(v))' $   
 for the subject-specific effects;  
$ \bPhij  = \left(\varphi_{j\nu}(t) \right)_{1\le \nu\le q,\, 1\le t \le T_j} $ and 
  $ \bd_j (v) = \left(d_{j1}(v),\ldots,d_{jq}(v) \right)' $ for the nuisance signals; 
and $ \bepsilonj(v) = \left( \varepsilon_j(v,1),\ldots, \varepsilon_j(v,T_j)\right)'$ for the noise. 
The symbol $\otimes$ denotes the Kronecker product.

The vector $\byj (v)$
has a multivariate normal distribution with mean and covariance  
\begin{equation}\label{mean and covariance of y_ij}
\begin{array}{l}
\displaystyle \boldsymbol{\mu}_j (v)=    \mathbf{X}_j\left(  \boldsymbol{\beta}(v) \otimes \boldsymbol{\gamma}(v) \right)
+ \boldsymbol{\Phi}_j \, \mathbf{d}_j(v), \medskip \\
\bVj(v) =    \bXj \big(\bD_{\xi}(v) \otimes \bI_K\big)\bT_\xi \bXj'
+  \bV_{\varepsilon jm} , 
\end{array}
\end{equation}
where $\bD_\xi(v) = \mathrm{diag}(\sigma^2_{\xi 1}(v),\ldots,\sigma^2_{\xi L}(v))$, 
$\bT_{\xi} = \mathrm{diag}(\bT_{\xi 1},\ldots,\bT_{\xi L}) $  is a block diagonal matrix with 
$\bT_{\xi l} = (\rho_{\xi l } (|k-k'| ) )_{1\le k,k' \le K} $, $\bI_K$ is the $K\times K$ identity matrix, 
and  $  \bV_{\varepsilon jm}$ is the covariance matrix of $\bepsilonj(v)$ for $v\in \mathcal{D}_m$. 
Here the first term of $\bVj(v)$ corresponds to between-subject variation, while the second term is the within-subject variation.

Table 1 summarizes all model notations. 

\begin{table}[ht]
\caption{Notations}
\begin{center}
\medskip
\begin{tabular}{ l l }
\hline 
$j$ & Subject index  \\
$k$ & Basis function index \\
$l$ & Condition index \\
$v$ & Voxel index \\
$n$ & Sample size \\
$K$ & Number of basis functions \\
 $L$ & Number of experimental conditions \\ 
$V$ & Number of voxels \\
$B_k$ & Basis function \\
$T_j$ & Number of brain scans for subject $j$ \\ 
$\byj(v) $& fMRI time course for subject $j$ at voxel $v$ \\
$\bXj $ & Design matrix for subject $j$ \\
 $\mathbf{X}_{jl} $ & Design matrix for subject $j$, condition $l$ \\
$\bbeta(v)$ & Amplitude parameters for the population HRF at voxel $v$ \\
$\bgamma(v) $ & Shape parameters for the population HRF at voxel $v$ \\
$ \bxij (v) $ & Deviation of subject $j$ from population HRF at voxel $v$ \\
$\bxijl(v)$  & Deviation of subject $j$ from population HRF for  condition $l$ and voxel $v$  \\
$ \bPhij  $ &Matrix of nuisance signals for subject $j$ \\
$ \bd_j (v)$ &nuisance coefficients for subject $j$ and voxel $v$\\
$ \bepsilonj(v) $ & Noise vector for subject $j$ at voxel $v$  \\
$\bD_\xi(v)$ & Variance coefficients of subject effects at voxel $v$ \\ 
$\bT_{\xi} $ & Correlation matrix for subject effects at voxel $v$ \\ 
$\bT_{\xi l} $ & Correlation matrix for subject effects for condition $l$ and voxel $v$ \\
 $  \bV_{\varepsilon jm}$ &  Covariance matrix of the noise for subject $j$ and parcel $m$ \\
 $\bVj(v)$ & Covariance matrix of $\byj(v) $ \smallskip \\
 \hline
\end{tabular}
\end{center}
\label{default}
\end{table}%

 \subsection{Estimation}
 \label{sec: estimation}
In this section we outline our procedure for estimating the parameters of our model. 
Our main objective is to estimate the population HRF at each voxel while taking into account the covariance structure of the data. 
We formulate this objective mathematically as a generalized, penalized, and constrained least squares problem.  
Since the corresponding objective function is non-convex, our procedure is only guaranteed to yield a local minimum.    
It is therefore important to select good starting values for the procedure, which we do by constructing a consistent pilot estimator of the HRF.  
We then provide consistent estimators of the data covariance parameters, after which we optimize the objective function to obtain the final HRF estimates. 
The entire procedure is performed in the following five steps, each described in detail in a subsequent subsection:

 \begin{enumerate}
 
\item For each voxel $v$, estimate the HRF parameters $ \gamma_{kl}(v) $ by Penalized Least Squares (PLS). 
This pilot estimation does not exploit the HRF shape assumption \eqref{simplification hrf} and does not  account for the covariance structure of the data. 

\item For each parcel $\mathcal{D}_m$, estimate the parameters $\sigma_{\varepsilon m}^2$ and $\boldsymbol{\theta}_{\varepsilon m}$ 
of the AR noise process $\varepsilon$ by solving the Yule-Walker equations associated with the predicted errors $\hatepsilonj(v)$. 
The $\hatepsilonj(v)$ are obtained from a least squares fit on the residuals of step 1. 
  
\item Estimate the temporal correlation parameters $\rho_{\xi l }(k)$ of the subject random effects  
by Maximum Likelihood (ML).   
The ML estimates are obtained separately on a small sample of voxels 
and pooled with a suitable statistic (e.g., trimmed mean or median).

\item For each voxel $v$,  estimate the between-subject variance $\sigma_{\xi}^2(v)$ by Variance Least Squares (VLS).

\item For each voxel $v$, estimate  $ \bbeta(v) $ and $ \bgamma(v)$ again 
using a generalized, penalized, and constrained least squares approach.

\end{enumerate}

\subsubsection*{Step 1: Pilot estimation of the HRF}
\label{sub: hrf and nuisance estimation}

For each  voxel $v$, the HRF scale and shape coefficients $\bbeta(v)$ and $\bgamma(v)$ 
are first estimated by penalized least squares (PLS):
\begin{equation}\label{pilot alpha_l}
\min_{\mathbf{h}, \mathbf{d}} \Bigg\{ \sum_{j=1}^n \big\|  \byj(v) -   \bXj
 \mathbf{h}  - \bPhij \bdj \big\|^2  + n\lambda_0 \,  \mathbf{h}' (\bI_L\otimes \bP) \mathbf{h} \Bigg\} \, ,
\end{equation}
where $\bh$ is a vector of length $KL$ that estimates the HRF coefficients $\gamma_{lk}(v)$ and  
$\bd = (\bd_1' ,\ldots ,\bd_n')'$ is a vector of length $nq$ that estimates  the nuisance signals. 
The matrix $\bP$ penalizes departures of $\bh$ from a linear space of ``reasonable" HRFs 
(e.g., the canonical HRF and its temporal derivative). 
More precisely, let $\boldsymbol{\Psi} $ be a matrix whose columns contain the coefficients of a few realistic HRFs 
  in the function basis $(B_k)_{1\le k \le K}$.
Then $\bP = \bI_K - \boldsymbol{\Psi}(\boldsymbol{\Psi}'\boldsymbol{\Psi})^{-1} \boldsymbol{\Psi}'$ is the projection on the orthogonal space of  $\boldsymbol{\Psi}$. The smoothing parameter
 $\lambda_0>0$  determines the tradeoff between fitting the data 
and closeness to $\boldsymbol{\Psi}$.
It can be selected manually or, for example, by $k$-fold cross-validation with the subjects randomly partitioned in $k$ subsamples. 

Note that the minimization problem \eqref{pilot alpha_l}  is unconstrained and does not rely on \eqref{simplification hrf}. 
Its solutions 
are $\hat{\bh}(v) = ( \sum_{j=1}^n \bXj ' \bRj \bXj +  n  \lambda_0 (\bI_L\otimes \bP)  )^{-1} \sum_{j=1}^n \bXj ' \bRj \byj (v) $ and 
$\hat{\bd}_j (v) =
(\bPhij'\bPhij)^{-1} \bPhij'
(\byj (v) - \bXj \hat{\bh}(v) ), \, 1\le j \le n$, 
where $\bRj = \mathbf{I}_T- \bPhij (\bPhij'\bPhij)^{-1}\bPhij' $ is the projection matrix on the orthogonal space of $\boldsymbol{\Phi}_j$. 
The estimator $\hat{\bh}(v)$  is consistent and asymptotically unbiased  for the $\gamma_{lk}(v)$
as the sample size $n\to\infty$ 
and the smoothing parameter $\lambda_0\to 0$. In  view of  \eqref{simplification hrf}, it follows that 
  $\hat{\bbeta}_{0}(v) =  (\|   \hat{\bh}_1(v) \|, \ldots, \|   \hat{\bh}_L(v) \| )'$ is a consistent estimator
of $\bbeta(v)$, where $\hat{\bh}(v) = (\hat{\bh}_1(v)',\ldots,\hat{\bh}_{L}(v)')'$
and the vectors $\hat{\bh}_l(v) , 1\le l \le L,$ have length $K$.
Similarly, the scaled average $\hat{\bgamma}_{0}(v) = \sum_{l=1}^L \hat{\bh}_l(v) /  \| \sum_{l=1}^L  \hat{\bh}_l(v) \| $ 
consistently estimates $\bgamma (v)$ when the latter is well-defined, i.e., when the voxel $v$ is activated by at least one experimental condition.

\subsubsection*{Step 2: Estimation of the noise structure}

We turn to the estimation of the noise parameters $\sigma^2_{\varepsilon m} $ and $\boldsymbol{\theta}_{\varepsilon m} $ in each parcel $\mathcal{D}_m, \, 1\le m \le M$.  
For each subject $ 1 \le j \le n $ and voxel $ v \in \mathcal{D}_m$, consider the residual vector $\brj(v) = \byj(v) - \bXj (\hat{\bbeta}_0(x)\otimes \hat{\bgamma}_0(v)) - \bPhij \hat{\mathbf{d}}(v) = \bRj (\byj(v) -  \bXj(\hat{\bbeta}_0(x)\otimes \hat{\bgamma}_0(v)) )$ resulting from step 1. Given \eqref{time course model} and the consistency of $\hat{\mathbf{h}} = \hat{\mathbf{h}}(v)$ in step 1, it holds that $\brj(v) \approx \bXj \bxij + \bepsilonj(v)$ for $n$ large enough and $\lambda_0$ small. 
The random effects $\bxij(v)$ and $\bepsilonj(v)$ can thus be predicted by least squares based on $\brj(v)$, yielding 
\begin{equation*}\label{prediction xi eps}
\hat{\bxi}_j(v) = (\bXj'\bXj)^{-1}  \bXj'\brj(v), \quad \hatepsilonj(v) = \brj(v) - \bXj \hat{\bxi}_j(v) . 
\end{equation*}
If the design matrix $\bXj$ is not full rank, the inverse $(\bXj'\bXj)^{-1}$ in the above formula 
is not defined and can be replaced by its pseudoinverse $ (\bXj'\bXj)^{+} $.
We then solve the Yule-Walker equations (see e.g., \cite[p. 239]{BrockwellDavis06}) associated with $\hatepsilonj(v)$, 
producing consistent estimates of  $\sigma^2_{\varepsilon m} $ and $\boldsymbol{\theta}_{\varepsilon m} $ for each subject $ 1\le j \le n$ and voxel $  v\in \mathcal{D}_m$.  
By taking the medians of these estimates across subjects and voxels, 
we obtain robust estimates $\hat{\sigma}^2_{\varepsilon m} $ and $\hat{\boldsymbol{\theta}}_{\varepsilon m} $.


\subsubsection*{Step 3: Estimation of the temporal dependence in subject-specific effects}
\label{sub: estim temp corr space var}

We estimate the temporal correlation parameters $ \rho_{\xi l }(k), \,1\le k \le K-1, \, 1\le l \le L,$ by Maximum Likelihood (ML).   
For computational efficiency,  ML estimates are separately produced at a small number of voxels and aggregated with a suitable statistic such as the median or trimmed mean. In practice, we propose to select a random sample of about 1000 voxels for the ML estimation. 
Following a common usage, we first perform a few iterations of the EM algorithm to provide good starting values for the optimization of the likelihood function.

In the rest of this section, we fix a voxel and omit ther index $v$ for conciseness. 
Writing   $ \boldsymbol{\sigma}_{\xi }^2 = ( \sigma_{\xi 1}^2,\ldots, \sigma_{\xi L}^2)' $ and 
$\boldsymbol{\rho}_{\xi}= (\rho_{\xi 1 }(1),\rho_{\xi 1 }(2),\ldots, \rho_{\xi L}(K-1))'$, 
we optimize the log-likelihood function (multiplied by $-2$) 
\begin{equation}\label{loglik}
\begin{split}
& L (\bbeta, \bgamma,\bd, \boldsymbol{\sigma}_{\xi }^2,  \boldsymbol{\rho}_{\xi },\sigma_{\varepsilon m}^2 , \boldsymbol{\theta}_{\varepsilon m}) \\
& \qquad = 
\sum_{j=1}^n \ln\left| \bVj
 \right| 
+    \sum_{j=1}^n \left(   \byj - \bXj  (\bbeta \otimes \bgamma) - \bPhij \bd_j \right)' \mathbf{V}_j^{-1} \left(   \byj - \bXj  (\bbeta \otimes \bgamma) - \bPhij \bd_j \right)
\end{split}
\end{equation}
with respect to   $\boldsymbol{\sigma}_{\xi }^2 $ and $ \boldsymbol{\rho}_{\xi }$ while fixing $\bbeta, \bgamma,\bd,\sigma_{\varepsilon m}^2$, and  
$\boldsymbol{\theta}_{\varepsilon m} $ to their previously estimated values. 
Note that although  we are only concerned here with the estimation of $ \boldsymbol{\rho}_{\xi }$, 
the likelihood must also be optimized with respect to $\boldsymbol{\sigma}_{\xi }^2 $. 
These variance parameters, which must be estimated at each voxel, will be assessed more efficiently in step 4. 
The implementation of the EM algorithm and likelihood optimization is described in Appendix \ref{app: temp dependence}. 
More details on the EM algorithm for linear mixed models can be found in  e.g., (\cite{Pawitan2001}, chap. 12).


 \subsubsection*{Step 4: Estimation of the between-subjects variance}
\label{sec: VLS}

For each voxel $v$, we  estimate the between-subjects variances $\boldsymbol{\sigma}_{\xi }^2 (v)$ 
by a variance least squares approach (e.g., \cite{Demidenko2004}, chap. 3) 
that consists of minimizing the distance between the residual covariance matrix 
and the theoretical covariance matrix:    
\begin{equation}\label{VLS objective}
\min_{ \boldsymbol{\sigma}_{\xi }^2 }\  \sum_j \Big\| \brj(v) \brj '(v) -  
 \bXj \big(\bD_{\xi}(v) \otimes \bI_K\big)\hat{\bT}_\xi \bXj'
-   \hat{\bV}_{\varepsilon jm} \Big\|^2_F
\end{equation}
subject to the constraint $\sigma_{\xi l}^2 \ge 0$ for  $1\le l \le L$. 
Recall that the residuals $\brj(v)$ are defined in step 2 of this section,  
$\bD_{\xi} = \mathrm{diag} ( \sigma_{\xi 1}^2,\ldots, \sigma_{\xi L}^2 )$, 
and $ \hat{\bV}_{\varepsilon jm}$ and $\hat{\bT}_\xi$ are the estimates of  $\bV_{\varepsilon jm}$ and $\bT_\xi$ obtained in steps 2-3. 
The notation $\| \bA \|_F $ stands for the  
 Frobenius norm $ \mathrm{tr}(\bA'\bA)^{1/2}$ of a matrix $\bA$.  
Problem \eqref{VLS objective} is a standard quadratic programming problem 
that expressed more simply as 
\begin{equation}\label{VLS reloaded}
\min_{ \boldsymbol{\sigma}_{\xi }^2 }  \Big\{   (\boldsymbol{\sigma}_{\xi }^2) ' \bA  \boldsymbol{\sigma}_{\xi }^2
- 2 \mathbf{b}'  \boldsymbol{\sigma}_{\xi }^2  \Big\} ,
\end{equation}
where $\bA$ is a $L\times L$ matrix with $(l,l')$ entry 
$\sum_{j=1}^n \mathrm{tr} \big( \bXjl  \hat{\bT}_{\xi l} \bXjl' \mathbf{X}_{jl'} \hat{\bT}_{\xi l'} \mathbf{X}_{jl'}' \big)$
and $\mathbf{b}$ is a vector of length $L$ with $l^{th}$ entry  
$\sum_{j=1}^n \big\{  \brj '(v) \bXjl  \hat{\bT}_{\xi l}  \bXjl'  \brj(v)
 -    \mathrm{tr} \big( \hat{\bT}_{\xi l} \bXjl ' 
 \hat{\bV}_{\varepsilon jm} \bXjl \big) \big\} $.
The solutions to \eqref{VLS objective}-\eqref{VLS reloaded} 
can be computed by various methods (e.g., \cite{Nocedal2006}, p. 449)
that are widely available in software packages. 


 \subsubsection*{Step 5: Generalized least squares estimation of the HRF}
\label{sec: GLS}

The pilot estimation of the HRF can be improved upon in two ways: (i) by accounting for the dependence structure of the BOLD signal, 
and (ii) by imposing the form \eqref{simplification hrf} to the HRF estimates. 
To integrate these features in the estimation, we use a penalized, constrained, generalized least squares approach. 
For a given voxel $v$, let $\hat{\bV}_j(v) =\bXj (\hat{\mathbf{D}}_{\xi}(v) \otimes \bI_K)    \hat{\bT}_\xi  \bXj'+
\hat{\mathbf{V}}_{\varepsilon jm}$ be the estimate of $\bVj(v)$ resulting from steps 1-4. 
We seek to solve 
\begin{equation}\label{constrained gls}
\min_{\bbeta ,\bgamma , \bd   } \Bigg\{
 \sum_{j=1}^n   \Big\| \byj(v)  - \bXj (\bbeta \otimes\bgamma )  -  \bPhij \bdj \Big\|_{\hat{\mathbf{V}}_j^{-1}(v)}^2 + n \lambda\, \bgamma'  \bP \bgamma \Bigg\}
\end{equation}
under the constraint $\| \bgamma \|^2 = 1$. 
Like in the pilot estimation, 
the nuisance parameter $\bd$ can be eliminated from \eqref{constrained gls}. 
To that intent, let $\btildeRj(v) = \mathbf{I}_T -  \bPhij ( \bPhij ' \hatVjinv(v) \bPhij )^{-1}  \bPhij '  \hatVjinv (v)$ 
be the projection matrix on the orthogonal space of $\bPhij$ in the metric $\hatVjinv (v)$. 
Then \eqref{constrained gls} is equivalent to $\min_{\bbeta ,\bgamma    } \big\{
 \sum_{j=1}^n   \big\| \btildeRj(v) \byj(v)  - \btildeRj(v) \bXj (\bbeta \otimes\bgamma )  \big\|_{\hatVjinv (v)}^2 + n \lambda\, \bgamma'  \bP \bgamma \big\}
$.  

Because of the tensor product $\bbeta \otimes \bgamma$ and the quadratic constraint $\|\bgamma\|^2=1$, 
problem \eqref{constrained gls} is nonlinear  and has no closed-form solutions. 
However,  \eqref{constrained gls}  is a separable least squares problem:  
 for a fixed $\bgamma$, solving \eqref{constrained gls} with respect to $\bbeta$ reduces to a generalized least squares problem that admits a closed-form solution. For a fixed $\bbeta$, solving \eqref{constrained gls} with respect to $\bgamma$ is a quadratically constrained quadratic program that 
 requires little more than a singular value decomposition. 
As a result, \eqref{constrained gls} can be efficiently solved in an iterative way. 
 
 For conciseness, we omit the index $v$ from notations in the remainder of the section. 
Let $ \bM =  \sum_{j=1}^n \bXj '  \btildeRj' \hatVjinv \btildeRj  \bXj$   and $\boldeta = \sum_{j=1}^n \bXj '   \hatVjinv \btildeRj \byj$.
The solutions $\hat{\bbeta}$ and $\hat{\bgamma}$ of  \eqref{constrained gls} are obtained by 
cycling through the following equations until convergence: 

\begin{flalign} 
\label{update beta gls}
 \hat{\bbeta} &  = 
\left[ \left( \bI_L \otimes \hat{\bgamma} \right)' 
\bM  \left( \bI_L \otimes \hat{\bgamma} \right) \right]^{-1} 
   \left( \bI_L \otimes \hat{\bgamma} \right)'  \boldeta \, , & \\ 
\label{Lagrange gls}
\hat{C} & = \underset{C}{\operatorname{arg\,min}}  \bigg\{ \boldeta'  \big(\hat{\bbeta} \otimes\mathbf{I}_K  \big)  \Big[ \big(\hat{\bbeta} \otimes \bI_K  \big)'  \bM   \big( \hat{\bbeta} \otimes \bI_K \big) \hspace*{-.5mm} + \hspace*{-.5mm} n  \lambda \bP \hspace*{-.5mm}+ \hspace*{-.5mm} C \hspace*{.3mm} \bI_K \Big]^{-1}  \big(\hat{\bbeta} \otimes \bI_K  \big)' \boldeta \hspace*{-.5mm} +\hspace*{-.5mm} C \bigg\}, &\\
\label{update gamma gls}
  \hat{\bgamma} & = \big[
   \big( \hat{\bbeta} \otimes \bI_K  \big)'\,  \bM  \, \big( \hat{\bbeta} \otimes \bI_K \big) +  n  \lambda \bP
    + \hat{C}\, \bI_K \big]^{-1} \, \big( \hat{\bbeta} \otimes \bI_K  \big)' \, \boldeta \, ,&
\end{flalign}  
with $\hat{\bgamma} $ initially set to  the pilot estimator $ \hat{\bgamma}_0 $. 
Equation \eqref{update beta gls} corresponds to the generalized least squares problem that updates $ \hat{\bbeta}$ for a given $  \hat{\bgamma}$. 
Equations \eqref{Lagrange gls}-\eqref{update gamma gls} correspond to the quadratically constrained quadratic program 
that  updates  $  \hat{\bgamma}$  for a given $ \hat{\bbeta}$ using the method of Lagrange multipliers. 
The Lagrange multiplier $\hat{C}$ is computed by performing the singular value decomposition 
of  $ (\hat{\bbeta} \otimes \bI_K  )' \, \bM \,   ( \hat{\bbeta} \otimes \bI_K )  +  n  \lambda \bP$ and numerically finding the root of a monotone function (see e.g., \cite[chap. 6]{Golub2013} for details). 



\subsection{Inference}
\label{sec: inference}

In this section we discuss the sampling distribution of the HRF estimates and illustrate how to perform inference on model parameters.
As before, we omit the voxel index  $v$ from notations for conciseness. 
Recall that for a given voxel, the HRF shape parameter $\bgamma$ is well defined only if 
 at least one condition induces an activation in the voxel. 
 Under this assumption,  for a sufficiently large sample size $n$ and sufficiently small smoothing parameter $\lambda$,   
 the sampling distributions of $\hat{\bbeta}$ and $\hat{\bgamma}$ 
 can be  respectively approximated 
 by $N(\bbeta ,  [  ( \bI_L \otimes  \bgamma )' \bM    ( \bI_L \otimes  \bgamma  )  ]^{-1} )$ and 
  $N(\bgamma,    [  (  \bbeta \otimes\bI_K   )'\bM  ( \bbeta \otimes \bI_K ) ]^{-1}  )$,  
  where the matrices $\hat{\bV}_j $  are replaced by the true covariances  $\bVj $ in $\bM$
  and where $N(\boldsymbol{\mu}, \boldsymbol{\Sigma})$ denotes the multivariate normal distribution 
with mean $\boldsymbol{\mu} $ and covariance matrix $\boldsymbol{\Sigma}$. 
Further, neglecting the uncertainty about $\bgamma$ 
(and about the covariance matrices $\mathbf{V}_j$) in the estimation of $\bbeta$ and vice-versa, 
we obtain 
\begin{equation}\label{eq: inference}
\begin{split}
\smallskip
\hat{\bbeta} & \approx N\left( \bbeta ,  \big[  \left( \bI_L \otimes  \hat{\bgamma} \right)' \bM  
  \left( \bI_L \otimes  \hat{\bgamma}  \right)  \big]^{-1} \right) , \\ 
\hat{\bgamma} &\approx  N\left( \bgamma,    \big[  \big(  \hat{\bbeta} \otimes\bI_K   \big)'\bM
  \big( \hat{\bbeta} \otimes \bI_K \big) \big]^{-1}  \right).
\end{split}
 \end{equation}
 The first result allows us to perform inference to detect activation, as is commonly performed in the GLM setting. The second  results allows us to test hypotheses regarding the shape of the HRF. For example, after computing the values of $\bgamma$ that correspond to the canonical HRF, we could test for deviations in the shape across the brain based on the fact 
 that $\big(\hat{\bgamma}-\bgamma \big)' \big[ \big( \hat{\bbeta} \otimes\bI_K   \big)'\bM \big( \hat{\bbeta} \otimes \bI_K \big) \big]^{-1}\,   \big(\hat{\bgamma}-\bgamma \big)$ follows a $\chi^2$ distribution with $K$ degrees of freedom.
A theoretical justification of \eqref{eq: inference} is provided in Appendix B.

Recall that the above inference procedures rely on the assumption that the voxel under study is activated in at least one condition 
(otherwise $\bgamma$ is not well defined). 
This assumption amounts to the fact that at least one of the unconstrained HRF coefficients $\gamma_{lk}$ is not zero.  
It can easily be tested using the pilot estimator $\hat{\mathbf{h}}$ of the $\gamma_{lk}$ defined in \eqref{pilot alpha_l}. 
More precisely, $\hat{\mathbf{h}}$ has a normal distribution with asymptotic bias  zero and asymptotic variance  
$( \sum_{j=1}^n \bXj ' \bRj \bXj   )^{-1} (\sum_{j=1}^n \bXj ' \bRj \bVj\bRj\bXj )
( \sum_{j=1}^n \bXj ' \bRj \bXj   )^{-1} $ for  large $n$ and small $\lambda_0$.
If the null hypothesis $\gamma_{lk}=0$ for all $k,l$ fails to be rejected, then no further inference should be performed for this voxel. 
Another sensible approach is to use  \eqref{eq: inference} first to test for activations in all voxels and then to infer the HRF shape in activated voxels.


\subsection{Simulations}

The basic framework for this simulation study is similar to the one found in Lindquist et al. (2008), where a number of different HRF modeling approaches were evaluated. Inside a static brain slice, with dimensions $51 \times 40$, a set of $25$ identically sized squares, with dimensions $4 \times 4$, were placed to represent active regions (see Fig. 1A). Within each square, we simulated BOLD fMRI signal based on different stimulus functions, which varied systematically across the squares in terms of onset and duration. From left to right the onset of activation varied from the first to the fifth TR. From top to bottom, the duration of activation varied from one to nine TRs in steps of two. To create the response, we convolved the stimulus function in each square with SPMs canonical HRF, using a modified nonlinear convolution that includes an exponential decay to account for refractory effects with stimulation across time, with the net result that the BOLD response saturates with sustained activity in a manner consistent with observed BOLD responses (Wager et al., 2005). This procedure gave rise to a set of $25$ distinct HRF shapes. Fig. 1B shows examples of the $5$ HRFs with no onset shift, which are representative of the remaining HRFs. 

In total we performed five simulation studies in order to evaluate the properties of the proposed model. Below follows a description of each. \\

\noindent {\it Simulation 1:} In this simulation the TR was assumed to be $1$s long and the inter-stimulus interval was set to $30$s. This activation pattern was repeated to simulate a total of $10$ epochs. To simulate a sample of subjects and group random-effects, we generated $15$ subject datasets, which consisted of the BOLD time series at each voxel plus white noise, creating a plausible effect size (Cohen's d$=0.5$) based on observed effect sizes in the visual and motor cortex (Wager et al., 2005). The value of $\sigma^2_{\varepsilon}$ was set to $3$. In addition, a random between-subject variation with a standard deviation of size one third of the within-subject variation was added to each subject's time course. \\

\noindent {\it Simulation 2:} The second simulated data set was constructed in precisely the same manner as outlined in Simulation 1, except here instead of using white noise to simulate within-subject variation we used an AR(1) noise process with $\theta_{\varepsilon} = 0.3$. \\

\noindent {\it Simulation 3:} The third simulated data set was constructed, using the exact same process with AR(1) noise described in Simulation 2, expect here instead of using SPMs canonical HRF we used a subject-specific HRF. These were randomly generated using 20 B-spline basis sets with weights drawn from a normal distribution with mean equal to the weights corresponding to the canonical HRF and standard deviation $0.1$. \\

\noindent {\it Simulation 4:} The fourth simulated data set was constructed in precisely the same manner as outlined in Simulation 1, except here we allowed for two separate conditions. For both conditions the inter-stimulus interval was set to $30$s and the activation pattern was repeated to simulate a total of $10$ epochs. However, the two conditions were interleaved to begin $15$s apart from one another. The $\beta$ value for the two conditions were set to $0.5$ and $1$, respectively. All other parameters were set according to the description of Simulation 1. \\

\noindent {\it Simulation 5:} The fifth simulated data set was constructed in precisely the same manner as outlined in Simulation 1, except here we used a fast event-related design. The inter-stimulus interval was randomized across each trial using a uniform distribution between $10$ and $20$s. All other parameters were set according to the description of Simulation 1. \\

For each of the five simulations, the basic data sets of dimensions $51 \times 40 \times 300$ were fit voxel-wise using both using the standard GLM/OLS approach and our proposed hierarchical approach. For Simulations 1-3 an event-related stimulus function with a single spike repeated every $30$s was used for fitting the models to the data set described above. For Simulation 4 this was supplemented by a second stimulus function with a single spike repeated every $30$s corresponding to the timing of the second condition. Finally, in Simulation 5 we used a stimulus function with a single spike repeated according to the outcomes of the randomization scheme outlined above. 

This implies that for each simulation the square in the upper left-hand corner of Fig. 1A is correctly specified for the standard GLM while the remaining squares have activation profiles that are mis-specified to various degrees. When fitting our method we used $20$ B-spline basis functions of order $6$. While our proposed method provides multi-subject estimates and a framework for direct inference on these parameters, the standard OLS (ordinary least squares) approach to multi-subject analysis in fMRI involves using a two-stage model. It begins by fitting individual regression coefficients for each subject using a standard GLM. Thereafter group estimates of the parameters are obtained by averaging across subjects and variance components are obtained by computing the variance of the estimates across subjects. Since this analysis is performed on the estimated regression coefficients, the variance will contain contributions from both the standard error of the estimates and the between-subject variance components. This method is the most popular method in the neuroimaging community for estimating the parameters of a mixed-effects model \cite{mumford2009simple}.

After estimation we performed population level inference to determine whether $\beta$ was significantly different from $0$. In each case we performed a one-sided test. In order to control for multiple comparisons we used an FDR-controlling procedure with $q=0.05$ \cite{Genovese}.


\subsection{Experimental Data}

All subjects ($n = 20$) provided informed consent in accord with the Declaration of Helsinki, and the Columbia University Institutional Review Board approved all procedures. Subjects were all right-handed, as assessed by the Edinburgh Handedness Inventory, and free of self-reported history of psychiatric and neurological disorders, excessive caffeine or nicotine use, and illicit drug use. They were pre-screened during an initial calibration session to ensure that stimuli were painful and that they could rate pain reliably ($r \ge .65$) between applied temperature and pain rating). During fMRI scanning, $48$ thermal stimuli, $12$ at each of $4$ temperatures, were delivered to the left forearm using a Peltier device (TSA-II, Medoc, Inc.) with an fMRI-compatible $1.5$ mm-diameter thermode. Temperatures were calibrated individually for each participant before scanning to be warm, mildly painful, moderately painful, and near tolerance. Heat stimuli were preceded by a $2$s warning cue and $6$s anticipation period, lasted $10$s in duration ($1.5$s ramp up/down, $7$s at peak), and were followed by a $30$s inter-trial interval (ITI). At a time $14$s into the ITI, participants were asked to rate the painfulness of the stimulus on an 8-point visual analogue scale using an fMRI-compatible trackball (Resonance Technologies, Inc.). 
Functional T2*-weighted EPI-BOLD images (TR = $2$s, $3.5 \times 3.5 \times 4$mm voxels, ascending interleaved acquisition) were collected during $6$ functional runs each consisting of $6$ minutes $8$s. Images were corrected for slice-timing acquisition delay and realigned to adjust for head motion using SPM5 software (http://www.fil.ion.ucl.ac.uk/spm/). A high-resolution anatomical image (T1-weighted spoiled-GRASS [SPGR] sequence, $1 \times 1 \times 1$mm voxels, TR $= 19$ms) was collected after the functional runs and coregistered to the mean functional image using a mutual information cost function (SPM5, with manual checking and adjustment of starting values to ensure satisfactory alignment for each participant), and was then segmented and warped to the Montreal Neurologic Institute template (avg152T1.nii) using SPM5's ``generative'' segmentation \cite{ashburner2005unified}. Warps were applied to functional images. Functional images were smoothed with a $6$ mm-FWHM Gaussian kernel, high-pass filtered with a $120$s ($.0083$ Hz) discrete cosine basis set (SPM5), and windsorized at $2$ standard deviations prior to analysis. 
Each of the models described above was fit to data from each voxel in a single axial slice (z $= -22$mm) that covered several pain-related regions of interest, including the anterior cingulate cortex. Separate HRFs were estimated for stimuli of each temperature, though we focus on the responses to the highest heat level in the results. 

We fit the data using our proposed hierarchical approach with $12$ B-spline basis functions of order $6$. As the fMRI data consisted of $6$ functional runs for each subject, we combined the results across runs in a manner corresponding to a fixed-effects model, assuming the same HRF for all runs for a given subject and voxel. As an alternative it would be possible to extend our model to allow $\beta$ to be a random effect across runs in a manner analogous to that outlined in \cite{BadilloVC13}. After estimation we performed population level inference to determine whether $\beta$ was significantly different from $0$. In each case we performed a one-sided test. In order to control for multiple comparisons we used an FDR-controlling procedure with $q=0.05$ \cite{Genovese}.


\section{RESULTS}

\subsection{Simulations}

Figs. 2-4 show results of the first three simulation studies. Each contains estimates of $\beta$, $\theta_{\varepsilon}$, $\sigma^2_{\varepsilon}$, the $t$-map, and the thresholded $t$-map, obtained using both the standard GLM/OLS approach and our hierarchical model. In each case, GLM/OLS (first row) gives reasonable results for delayed onsets within $3$s and widths up to $3$s, corresponding to squares in the upper left-hand corner. However, its performance worsens dramatically as onset and duration increase. As pointed out in Lindquist et al. (2008) this is natural as the GLM is correctly specified for the square in the upper left-hand corner, but not well equipped to handle large deviations from this model. Interestingly, in Fig. 2 we see that this model misspecification gives rise to a positive autocorrelation in several of the most severely effected squares, even though we used white noise in that particular simulation. 

Almost all of these problems are solved using our hierarchical model. Clearly, irrespective of the shape of the underlying HRF we were able to efficiently recover both $\beta$ and the variance component in each square. Clearly, these improved estimations lead to significantly improved sensitivity and specificity in the population-level hypothesis tests. 

The results of Simulation 4 are shown in Fig. 5. Here the first column contains the t-map corresponding to the test of whether the contrast, $\beta_1 - \beta_2$, between the parameters of the two conditions was significantly different from zero, while the second column shows the analogous thresholded t-maps. Again, our hierarchical model is able to detect significant regions with both high sensitivity and specificity. In contrast, the GLM/OLS approach performs poorly, even in the upper-left hand corner where it is expected to be optimal.

Fig. 6 shows results for Simulation 5 where we used a rapid event-related task. The first column contains the t-map corresponding to testing whether the parameter $\beta$ was significantly different from zero, and the second column corresponds to the thresholded t-maps. The GLM/OLS approach performs somewhat better than in the other simulations, perhaps due to the increased number of trials. However, our hierarchical model is again able to detect significant regions with both a high degree of sensitivity and specificity. 

To illustrate further, we see in Fig. 7 the results of $100$ replications of the first simulation. Fig. 7A shows the portion of times the standard GLM/OLS approach gave significant results in each voxel across the simulated brain. Fig. 7B shows the same results for our hierarchical model. Clearly, our hierarchical model is able to consistently detect truly activated regions, while avoiding spurious findings. Not surprisingly, the GLM/OLS approach only performs well in squares where it is correctly defined. Fig 7C shows the portion of times the estimated HRF significantly deviated from SPMs canonical form ($\alpha <0.05$). This test was performed by first computing values of $\gamma$ corresponding to the canonical HRF and thereafter using the $\chi^2$ test described in Section \ref{sec: inference}. The results show, again not surprisingly, that the HRF in voxels in the lower right-hand corner of the brain deviate significantly from the canonical form. Interestingly, these voxels corresponds closely to voxels that were erroneously deemed non-active in Fig. 7A. This leads us to believe that our approach could have an alternative use as a diagnostic tool for assessing the performance of standard GLM analyses. 

Fig 8 shows the methods ability to recover the time-to-peak and width of the underlying HRF used to generate the data. The left hand column shows the true values and the right hand columns the mean estimated values across the $100$ replications. Clearly we are able to extremely accurately capture the true value of the time-to-peak. However, it appears that the estimates of the width, while close are somewhat confounded by the changes in onset, with the best results occurring when there are no onset shifts present.


\subsection{Experimental Data}

The results of the pain experiment are shown in Fig. 9 for a single axial slice (z $= -22$mm). The location of the slice used and an illustration of key areas of interest are shown in Fig. 9A. The highlighted areas are the rostral dorsal anterior cingulate cortex (rdACC) and the secondary somatosensory cortex (S2); two brain regions known from previous work to be involved in the processing of pain intensity \cite{ferretti2003functional, peyron2000functional}. Activation in S2 is thought to be related to sensory-discriminant aspects of pain-processing, while rdACC has been shown to be related to expectancy \cite{atlas2012dissociable}.

In Fig. 10 we see examples of the estimated HRFs, obtained using our approach, from voxels chosen because they lay in the center of the rdACC and S2, respectively. These results include both the subject-specific and group-level estimates. The shapes of both group-level HRFs are significantly different from the canonical HRF ($p <0.01$) using the $\chi^2$ test. Note that the shapes of the HRFs are also quite different from one another. For rdACCC, the width of the response is significantly wider from what we would expect from the canonical HRF. While for S2, the onset is significantly delayed. However, interestingly both HRFs appear to reach their peak at roughly the same time point following activation.

Due to the apparent variability in the HRF across the slice, it would be problematic to analyze this data set using a canonical HRF or one that uses a constrained basis set. In fact, the GLM showed no activation in either rdACC or S2. However, the flexibility of our approach allows for large deviations in the shape of the HRF across voxels. In Fig. 9B we see an activation map obtained using our method. In particular, note that there is significant activation in S2 in response to the noxious stimulation. In previous analysis of this data set \cite{lindquist8}, activations in this region were particularly difficult to detect using standard GLM methods (e.g., the canonical HRF plus its temporal and dispersion derivatives, or the finite impulse response (FIR) basis set). The inverse-logit (IL) model was the only approach that showed activation in S2 contralateral to noxious stimulation.  The activation shown here is extremely robust in comparison.


\section{DISCUSSION}

In this paper we introduce a new approach towards the simultaneous detection of activation and the estimation of the HRF for multi-subject fMRI. The suggested approach circumvents a number of shortcomings in the standard approach for performing group analysis. In these approaches there is often a tension between flexible modeling of the HRF at the first level and straightforward inference in the second level. For example, if multiple basis sets are used in the first-level GLM, then it is difficult to determine an approriate contrast to bring forward to the second-level. Often, researchers use only a subset of the basis functions and therefore potentially ignore important information contained in those left behind.  The proposed approach allows the shape of the hemodynamic response function to vary across regions and subjects, as when using basis sets, while still providing a straightforward way to estimate population-level activation using all the information from the first level analysis. 

An additional benefit of our model is that the suggested inferential framework not only provides a means for performing the standard tests for determining whether a voxel is significantly active, but also allows one to test whether the estimated HRF deviates from some canonical shape (e.g., SPMs canonical HRF). This type of inference has been under-utilized in the field, but we feel it can be extremely useful for diagnostic purposes, as it can help identify regions that would normally not be deemed active when using a canonical HRF.
 
In addition, our method could prove useful in situations when the standard HRF is either ill-fitting, such as in studies of young or elderly populations, or when the exact onset time or width of activation is unknown and added flexibility is needed to properly fit the data.  An example of the latter is the thermal pain data presented in this paper. In previous studies \cite{lindquist8}, activations in S2 were particularly difficult to detect using either the canonical HRF plus its temporal and dispersion derivatives, or the finite impulse response (FIR) basis set. However, using our approach, the activation was extremely robust in comparison.

To date the method has only been implemented in a voxel-size manner. Hence, we make the common, but rather implausible, assumption of independence between voxels with regards to both the HRF shape and amplitude. However, we do allow for the possibility of a parcel-specific noise structure. We are currently in the process of extending the method to also estimate spatial dependences in the data. For that purpose two types of approaches can be considered: one consists in spatially regularizing the estimation in local neighborhoods; the other is to  integrate a functional parcellation of the brain. In the latter case, one can either resort to an existing atlas (see e.g. \cite{Karahano2013}) or define a data-driven parcellation (see e.g., \cite{Chaari2012hemo} for such a parcellation at the subject level).
Though to make the manuscript more manageable we decided to limit our discussion of this issue until a later time.

The presented simulations and data set are identical to those used in a  previous study \cite{lindquist8}, where we evaluated the performance of seven different HRF models. These included: SPMs canonical HRF; the canonical HRF plus its temporal derivative; the canonical HRF plus its temporal and dispersion derivatives; the finite impulse response (FIR) basis set; a regularized version of the FIR model (denoted the smooth FIR); a nonlinear model with the same functional form as the canonical HRF but with 6 variable parameters; and the inverse logit (IL) model. The results of that work showed that it was surprisingly difficult to accurately recover true task-evoked changes in BOLD signal and that there were substantial differences among models in terms of power, bias and parameter confusability. While the derivative models were accurate for very short shifts in latency they became progressively less accurate as the shift increased. The IL model and the smooth FIR model showed the least amount of  biases, and the IL model showed by far the least amount of confusability of all the models that were examined. Both these methods were clearly able to handle even large amounts of model misspecification and uncertainty about the exact timing of the onset and duration of activation.

The suggested model clearly outperformed each of the 7 other models in the same battery of tests. For space purposes we only present the canonical HRF for comparison purposes, and we encourage interested readers to 
look back at \cite{lindquist8}, for more results. The hierarchical model was found to have a superb balance of sensitivity and specificity that none of the other models was able to obtain. In addition, in the previous work the IL model was the only model that showed significant activation in S2 contralateral to noxious stimulation. Here the proposed model shows extremely robust signal in this region. For these reasons we believe the proposed model is a useful approach towards effectively modeling multi-subject fMRI data.

A Matlab implementation of the proposed methodology is available upon request. 
The code runs in a reasonable time for fMRI data sets of moderate size. 
By optimizing the code and running it in parallel on voxels and/or on subjects, our methodology can scale up to large fMRI data. 
Specifically, step 1 of the estimation algorithm 
can be solved in closed form and requires few matrix multiplications. 
Step 2 is also very fast due to the definition of the noise parameters at the parcel level 
and to the computational efficiency of the Yule-Walker equations. Step
3 (maximum likelihood/EM algorithm) would be very slow if it was run
for each voxel but is in reality only carried for a small number of voxels. 
Step 4 consists in a large number of standard quadratic programming problems that can be solved quickly and in parallel. 
Step 5 is arguably the slowest part of the estimation procedure because of the necessity to compute inverse data covariance matrices. 
However, linear algebra tricks can reduce the dimension of the matrices to be inverted from the number of scans ($T_j$)
to the number of regressors ($KL$).  


\section*{Acknowledgements}
We thank two anonymous reviewers for remarks that helped us improve the quality of this paper, and Tor Wager for supplying the data.
This research was partially supported by NIH grant R01EB016061.

\bibliographystyle{apalike}
\bibliography{fmri}

\begin{thebibliography}{}

\bibitem[Aguirre et~al., 1998]{aguirre}
Aguirre, G.~K., Zarahn, E., and D'Esposito, M. (1998).
\newblock The variability of human, {BOLD} hemodynamic responses.
\newblock {\em NeuroImage}, 8(4):360--369.

\bibitem[Ashburner and Friston, 2005]{ashburner2005unified}
Ashburner, J. and Friston, K.~J. (2005).
\newblock Unified segmentation.
\newblock {\em Neuroimage}, 26(3):839--851.

\bibitem[Atlas et~al., 2012]{atlas2012dissociable}
Atlas, L.~Y., Whittington, R.~A., Lindquist, M.~A., Wielgosz, J., Sonty, N.,
  and Wager, T.~D. (2012).
\newblock Dissociable influences of opiates and expectations on pain.
\newblock {\em The Journal of Neuroscience}, 32(23):8053--8064.

\bibitem[Badillo et~al., 2013a]{Badillo2013}
Badillo, S., Vincent, T., and Ciuciu, P. (2013a).
\newblock Group-level impacts of within- and between-subject hemodynamic
  variability in {fMRI}.
\newblock {\em NeuroImage}, 82(0):433 -- 448.

\bibitem[Badillo et~al., 2013b]{BadilloVC13}
Badillo, S., Vincent, T., and Ciuciu, P. (2013b).
\newblock Multi-session extension of the joint-detection framework in {fMRI}.
\newblock In {\em ISBI}, pages 1512--1515.

\bibitem[Brockwell and Davis, 2006]{BrockwellDavis06}
Brockwell, P.~J. and Davis, R.~A. (2006).
\newblock {\em Time series: theory and methods}.
\newblock Springer Series in Statistics. Springer, New York.
\newblock Reprint of the second (1991) edition.

\bibitem[Calhoun et~al., 2004]{Calhoun04}
Calhoun, V.~D., Stevens, M.~C., Pearlson, G.~D., and Kiehl, K.~A. (2004).
\newblock {fMRI} analysis with the general linear model: removal of
  latency-induced amplitude bias by incorporation of hemodynamic derivative
  terms.
\newblock {\em NeuroImage}, 22:252--257.

\bibitem[Chaari et~al., 2012]{Chaari2012hemo}
Chaari, L., Forbes, F., Vincent, T., and Ciuciu, P. (2012).
\newblock Hemodynamic-informed parcellation of fmri data in a joint detection
  estimation framework.
\newblock In Ayache, N., Delingette, H., Golland, P., and Mori, K., editors,
  {\em Medical Image Computing and Computer-Assisted Intervention - MICCAI
  2012}, volume 7512 of {\em Lecture Notes in Computer Science}, pages
  180--188. Springer Berlin Heidelberg.

\bibitem[Chaari et~al., 2013]{chaari2012fast}
Chaari, L., Vincent, T., Forbes, F., Dojat, M., and Ciuciu, P. (2013).
\newblock Fast joint detection-estimation of evoked brain activity in
  event-related {fMRI} using a variational approach.
\newblock {\em IEEE Transactions on Medical Imaging}, 32(5):821--837.

\bibitem[Demidenko, 2004]{Demidenko2004}
Demidenko, E. (2004).
\newblock {\em Mixed models}.
\newblock Wiley Series in Probability and Statistics. Wiley-Interscience [John
  Wiley \& Sons], Hoboken, NJ.
\newblock Theory and applications.

\bibitem[Ferretti et~al., 2003]{ferretti2003functional}
Ferretti, A., Babiloni, C., Gratta, C.~D., Caulo, M., Tartaro, A., Bonomo, L.,
  Rossini, P.~M., and Romani, G.~L. (2003).
\newblock Functional topography of the secondary somatosensory cortex for
  nonpainful and painful stimuli: an {fMRI} study.
\newblock {\em Neuroimage}, 20(3):1625--1638.

\bibitem[Friston et~al., 1998]{friston1998event}
Friston, K., Fletcher, P., Josephs, O., Holmes, A., Rugg, M., and Turner, R.
  (1998).
\newblock Event-related {fMRI}: characterizing differential responses.
\newblock {\em Neuroimage}, 7(1):30--40.

\bibitem[Genovese et~al., 2002]{Genovese}
Genovese, C., Lazar, N., and Nichols, T. (2002).
\newblock Thresholding of statistical maps in functional neuroimaging using the
  false discovery rate.
\newblock {\em NeuroImage}, 15:870--878.

\bibitem[Genovese, 2000]{Genovese2000}
Genovese, C.~R. (2000).
\newblock A bayesian time-course model for functional magnetic resonance
  imaging data.
\newblock {\em Journal of the American Statistical Association},
  95(451):691--703.

\bibitem[Glover, 1999]{glover1999deconvolution}
Glover, G.~H. (1999).
\newblock Deconvolution of impulse response in event-related {BOLD} {fMRI}.
\newblock {\em NeuroImage}, 9(4):416--429.

\bibitem[Golub and Van~Loan, 2013]{Golub2013}
Golub, G.~H. and Van~Loan, C.~F. (2013).
\newblock {\em Matrix computations}.
\newblock Johns Hopkins Studies in the Mathematical Sciences. Johns Hopkins
  University Press, Baltimore, MD, fourth edition.

\bibitem[Goutte et~al., 2000]{goutte}
Goutte, C., Nielsen, F.~A., and Hansen, L.~K. (2000).
\newblock Modeling the haemodynamic response in {fMRI} using smooth {FIR}
  filters.
\newblock {\em IEEE Transactions on Medical Imaging}, 19:1188--1201.

\bibitem[Handwerker et~al., 2004]{Handwerker2004}
Handwerker, D.~A., Ollinger, J.~M., and D'Esposito, M. (2004).
\newblock Variation of {BOLD} hemodynamic responses across subjects and brain
  regions and their effects on statistical analyses.
\newblock {\em NeuroImage}, 21(4):1639 -- 1651.

\bibitem[Karahano\v{g}lu et~al., 2013]{Karahano2013}
Karahano\v{g}lu, F.~I., Caballero-Gaudes, C., Lazeyras, F., and Ville, D. V.~D.
  (2013).
\newblock Total activation: fmri deconvolution through spatio-temporal
  regularization.
\newblock {\em NeuroImage}, 73(0):121 -- 134.

\bibitem[Kiehl and Liddle, 2001]{kiehl2001event}
Kiehl, K.~A. and Liddle, P.~F. (2001).
\newblock An event-related functional magnetic resonance imaging study of an
  auditory oddball task in schizophrenia.
\newblock {\em Schizophrenia Research}, 48(2):159--171.

\bibitem[Liao et~al., 2002]{Liao02}
Liao, C., Worsley, K.~J., Poline, J.-B., Duncan, G.~H., and Evans, A.~C.
  (2002).
\newblock Estimating the delay of the response in {fMRI} data.
\newblock {\em NeuroImage}, 16:593--606.

\bibitem[Lindquist et~al., 2009]{lindquist8}
Lindquist, M., Loh, J., Atlas, L., and Wager, T. (2009).
\newblock Modeling the hemodynamic response function in {fMRI}: Efficiency,
  bias and mis-modeling.
\newblock {\em NeuroImage}, 45(1, Supplement 1):S187 -- S198.

\bibitem[Lindquist, 2008]{lindquist2008statistical}
Lindquist, M.~A. (2008).
\newblock The statistical analysis of {fMRI} data.
\newblock {\em Statistical Science}, 23(4):439--464.

\bibitem[Lindquist and Wager, 2007]{lindquist6}
Lindquist, M.~A. and Wager, T.~D. (2007).
\newblock Validity and power in hemodynamic response modeling: A comparison
  study and a new approach.
\newblock {\em Human Brain Mapping}, 28:764--784.

\bibitem[Makni et~al., 2005]{makni2005joint}
Makni, S., Ciuciu, P., Idier, J., and Poline, J.-B. (2005).
\newblock Joint detection-estimation of brain activity in functional {MRI}: a
  multichannel deconvolution solution.
\newblock {\em IEEE Transactions on Signal Processing}, 53(9):3488--3502.

\bibitem[Makni et~al., 2008]{makni2008fully}
Makni, S., Idier, J., Vincent, T., Thirion, B., Dehaene-Lambertz, G., and
  Ciuciu, P. (2008).
\newblock A fully bayesian approach to the parcel-based detection-estimation of
  brain activity in {fMRI}.
\newblock {\em Neuroimage}, 41(3):941--969.

\bibitem[Mumford and Nichols, 2009]{mumford2009simple}
Mumford, J.~A. and Nichols, T. (2009).
\newblock Simple group {fMRI} modeling and inference.
\newblock {\em Neuroimage}, 47(4):1469--1475.

\bibitem[Nocedal and Wright, 2006]{Nocedal2006}
Nocedal, J. and Wright, S.~J. (2006).
\newblock {\em Numerical optimization}.
\newblock Springer Series in Operations Research and Financial Engineering.
  Springer, New York, second edition.

\bibitem[Pawitan, 2001]{Pawitan2001}
Pawitan, Y. (2001).
\newblock {\em In all likelihood: statistical modelling and inference using
  likelihood}.
\newblock Oxford science publications. Clarendon press, Oxford.

\bibitem[Peyron et~al., 2000]{peyron2000functional}
Peyron, R., Laurent, B., and Garcia-Larrea, L. (2000).
\newblock Functional imaging of brain responses to pain. {A} review and
  meta-analysis (2000).
\newblock {\em Neurophysiologie Clinique/Clinical Neurophysiology},
  30(5):263--288.

\bibitem[Poldrack et~al., 2011]{poldrack2011handbook}
Poldrack, R.~A., Mumford, J.~A., and Nichols, T.~E. (2011).
\newblock {\em Handbook of functional MRI data analysis}.
\newblock Cambridge University Press.

\bibitem[Riera et~al., 2004]{riera}
Riera, J.~J., Watanabe, J., Kazuki, I., Naoki, M., Aubert, E., Ozaki, T., and
  Kawashima, R. (2004).
\newblock A state-space model of the hemodynamic approach: nonlinear filtering
  of bold signals.
\newblock {\em NeuroImage}, 21:547--567.

\bibitem[Sanyal and Ferreira, 2012]{sanyal2012bayesian}
Sanyal, N. and Ferreira, M.~A. (2012).
\newblock Bayesian hierarchical multi-subject multiscale analysis of functional
  mri data.
\newblock {\em Neuroimage}, 63(3):1519--1531.

\bibitem[Schacter et~al., 1997]{Schacter}
Schacter, D.~L., Buckner, R.~L., Koutstaal, W., Dale, A.~M., and Rosen, B.~R.
  (1997).
\newblock Rectangular confidence regions for the means of multivariate normal
  distributions.
\newblock {\em Late onset of anterior prefrontal activity during true and false
  recognition: an event-related {fMRI} study}, 6:259--269.

\bibitem[Vincent et~al., 2010]{vincent2010spatially}
Vincent, T., Risser, L., and Ciuciu, P. (2010).
\newblock Spatially adaptive mixture modeling for analysis of fmri time series.
\newblock {\em Medical Imaging, IEEE Transactions on}, 29(4):1059--1074.

\bibitem[Woolrich et~al., 2004]{woolrich}
Woolrich, M.~W., Behrens, T.~E., and Smith, S.~M. (2004).
\newblock Constrained linear basis sets for {HRF} modelling using variational
  {B}ayes.
\newblock {\em NeuroImage}, 21(4):1748--1761.

\bibitem[Worsley and Friston, 1995]{worsley}
Worsley, K.~J. and Friston, K.~J. (1995).
\newblock Analysis of {fMRI} time-series revisited-again.
\newblock {\em NeuroImage}, 2:173--181.

\bibitem[Zarahn, 2002]{zarahn2002using}
Zarahn, E. (2002).
\newblock Using larger dimensional signal subspaces to increase sensitivity in
  {fMRI} time series analyses.
\newblock {\em Human brain mapping}, 17(1):13--16.

\bibitem[Zhang et~al., 2013]{Zhang2013}
Zhang, T., Li, F., Beckes, L., and Coan, J.~A. (2013).
\newblock A semi-parametric model of the hemodynamic response for multi-subject
  fmri data.
\newblock {\em NeuroImage}, 75(0):136 -- 145.

\end{thebibliography}

\newpage

\appendix 

{\Large \bf Appendix}

\section{Estimation of the dependence in subject effects}
\label{app: temp dependence}

\subsection{EM algorithm} 

We fix a voxel and omit the index $v$ from notations. 
Since the random effects $\bxij $ are independent,  identically distributed, and stationary, it holds that $\sigma_{\xi l}^2 = \mathbb{E}(\| \bxi_{jl}\|^2 )/K$ for each subject ($1\le j \le n$) and condition ($1\le l \le L$). Hence, we first estimate $\sigma_{\xi l}^2$ by the empirical average $\sigma_{\xi l}^{2(0)} =  (1/n) \sum_{j=1}^n  \|\hat{\bxi}_{jl} \|^2 / K $. We initially assume working independence  for the $\bxijl $ so that $\rho_{\xi l}^{(0)}(k) = \delta_{0k}$ for each lag ($1\le k \le K-1$) and condition. 
At the $(r+1)^{th}$ iteration ($r\ge 0$) of the EM algorithm, the E-step computes the conditional expectation of 
 (minus twice the logarithm of) the complete likelihood,  i.e., the likelihood of the augmented data $(\by_1,\ldots,\by_n,\bxi_1,\ldots, \bxi_n)$.  
The conditioning variables are $\by_1,\ldots,\by_n $, the current estimators $ \boldsymbol{\sigma}_{\xi}^{2(r)},\boldsymbol{\rho}_{\xi}^{(r)}$, 
and $\hat{\bbeta}_0, \hat{\bgamma}_0,  \hat{\sigma}_{\varepsilon m}^2 , \hat{\boldsymbol{\theta}}_{\varepsilon m } $.
Up to constant terms, the conditional expectation is 
\begin{equation}\label{EM E step}
\begin{split}
\hspace*{-3mm}Q \big(\boldsymbol{\sigma}_{\xi}^2,\boldsymbol{\rho}_{\xi} \big| \boldsymbol{\sigma}_{\xi}^{2(r)},\boldsymbol{\rho}_{\xi}^{(r)}\big) 
&= 
nK  \ln \left| \bD_{\xi} \right|  + n \ln  \left| \bT_{\xi }\right|  \\
&\quad + \sum_{j=1}^n  \bxi_j^{(r)'}   \bT_{\xi }^{-1}\big( \bD_{\xi}^{-1}\otimes \bI_K\big)  \bxi_j^{(r)} 
+ \sum_{j=1}^n   \mathrm{tr} \big\{    \bT_{\xi }^{-1}\big( \bD_{\xi}^{-1}\otimes \bI_K\big) \mathbf{B}_j^{(r)} \big\} ,
\end{split}
\end{equation}
where 
$ \mathbf{B}_j^{(r)}=\big[ \bXj' \hat{\bV}_{\varepsilon jm}^{-1} \bXj + \big(\bT_{\xi}^{(r)}\big)^{-1}
 \big(( \bD_{\xi}^{(r)})^{-1} \otimes \bI_K\big)  \big]^{-1} $, 
$ \bxi^{(r)}_j =\mathbf{B}_j^{(r)} \bXj' \hat{\bV}_{\varepsilon jm}^{-1} \,\brj $ is the predicted random effect for the $j^{th}$ subject, 
and $\brj $ is the residual vector defined in step 2 of section \ref{sec: estimation}.

For $1\le l \le L$, the derivative of $Q$ with respect to  $\sigma_{\xi l}^2$ is 
\begin{equation}\label{dQ-dsigma2}
\frac{\partial Q}{\partial \sigma_{\xi l}^2} = \frac{nK}{ \sigma_{\xi l}^2} 
- \frac{1}{\sigma_{\xi l}^4} \sum_{j=1}^n  \bxi_{jl}^{(r)'}   \bT_{\xi l}^{-1} \bxi_{jl}^{(r)}
- \frac{1}{\sigma_{\xi l}^4}  \sum_{j=1}^n   \mathrm{tr} \big\{    \bT_{\xi l }^{-1} \,
\mathbf{B}_{jll}^{(r)}\big\},
\end{equation}
 where the matrix $ \mathbf{B}_{j}^{(r)} $ has been partitioned in blocks $ \mathbf{B}_{jl l'}^{(r)} , 1\le l,l'\le L,$ of size $K\times K$.

 Writing $ \mathbf{C}_l^{(r)} =  \sum_{j=1}^n \big( \bxi_{jl}^{(r)} \bxi_{jl}^{(r)' } + \mathbf{B}_{jll}^{(r)} \big)$ and equating \eqref{dQ-dsigma2}  with zero,  we get 
\begin{align}\label{sigma xi EM}
 \sigma_{\xi l}^2 & = 
 \frac{1}{nK}\bigg( \sum_{j=1}^n  \bxi_{jl}^{(r)'}   \bT_{\xi l}^{-1} \bxi_{jl}^{(r)}+ \sum_{j=1}^n   \mathrm{tr} \big\{    \bT_{\xi }^{-1}
 \mathbf{B}_{jll}^{(r)} \big\}\bigg)\nonumber \\
& =  \frac{1}{nK}\,
 \mathrm{tr} \big( \bT_{\xi l}^{-1}\mathbf{C}_l^{(r)} \big).
\end{align}

After plugging \eqref{sigma xi EM} in \eqref{EM E step}, 
the variance-profile function $Q_p$ to be optimized in the M step of the algorithm is  
\begin{align}\label{profiled Q}
Q_p\big(\boldsymbol{\rho}_{\xi} \big| \boldsymbol{\sigma}_{\xi}^{2(r)},\boldsymbol{\rho}_{\xi}^{(r)} \big) &= nK  \ln \left| \bD_{\xi} \right|  + n \ln  \left| \bT_{\xi }\right|  + n \nonumber \\
& =  nK \sum_{l=1}^L \ln \big( \mathrm{tr}  \big( \bT_{\xi l}^{-1}\mathbf{C}_l^{(r)} \big) \big) - nK \ln(nK) +  
n\sum_{l=1}^L  \ln  \left| \bT_{\xi l}\right| + n .
\end{align}

For computational speed, we may use any suitable gradient-based optimization method.
Let $\bD_k$ be the $K\times K$ matrix whose 
$(i,j)$ entry  is 1 if $|i-j|=k$ and 0 otherwise. 
Then $\bT_{\xi l} = \sum_{k=0}^{K-1} \rho_{\xi l}(k) \bD_k$ and the gradient of $Q_p$ is 
\begin{equation}\label{gradient profiled Q}
\frac{\partial Q_p}{\partial \rho_{\xi l }(k)} =  -
nK \, \frac{ \mathrm{tr}  \big(     \bT_{\xi l}^{-1} \bD_{k}   \bT_{\xi l}^{-1} \mathbf{C}_l^{(r)} \big) } { \mathrm{tr}  \big( \bT_{\xi l}^{-1}\mathbf{C}_l^{(r)} \big)}
+ n\,  \mathrm{tr}  \big( \bT_{\xi l}^{-1} \bD_{k} \big) 
\end{equation}
for $1\le k \le K-1$ and $1\le l \le L$. 
The box constraints  $-1\le \rho_{\xi l}(k) \le 1$ 
and positive definiteness of $\bT_{\xi l}$ are enforced during the optimization.
The  updated variance estimator $\sigma_{\xi l}^{2(r+1)}$ 
 is obtained by plugging $\rho_{\xi}^{(r+1)}$ in \eqref{sigma xi EM}.
Writing  $\bGamma^{(r)} = \big(\mathrm{vec}(\bT_{\xi 1}^{-1}\mathbf{C}^{(r)}_1 \bT_{\xi 1}^{-1}),\ldots, \mathrm{vec}(\bT_{\xi L}^{-1} \mathbf{C}^{(r)}_L \bT_{\xi L}^{-1})\big)$, 
 $\bD = \big(\mathrm{vec}(\bD_1),\ldots,\mathrm{vec}(\bD_{K-1})\big) $, $\mathbf{t}^{(r)}=\big( \mathrm{tr}  \big( \bT_{\xi 1}^{-1}\mathbf{C}_1^{(r)} \big),\ldots,  \mathrm{tr}  \big( \bT_{\xi L}^{-1}\mathbf{C}_L^{(r)} \big)\big)'$, 
 and $\mathbf{S} = \big( \mathrm{vec} (\bT_{\xi 1}^{-1}),\break \ldots, \mathrm{vec} (\bT_{\xi L}^{-1})\big)$, 
 the gradient \eqref{gradient profiled Q} 
can be  compactly written as 
\[
\frac{\partial Q_p}{ \partial \boldsymbol{\rho}_{\xi} }= - nK \, \frac{\mathrm{vec} \big( \bD' \bGamma^{(r)}  \big) }{ \mathbf{t}^{(r)} \otimes \mathbf{1}_{ K-1}     }
 +n\, \mathrm{vec} \big( \bD' \mathbf{S}  \big)\, , \] 
where the division is taken element-wise and  $ \mathbf{1}_{ K-1}$ is a vector containing $(K-1)$ ones.

\subsection{Maximum Likelihood Estimation} 

For a given voxel $v$, 
 the partial derivatives of the likelihood function \eqref{loglik} are  
\begin{align}
\frac{\partial L}{\partial \sigma_{\xi l}^2 } & = \sum_{j=1}^n 
\mathrm{tr}\big( \bXjl' \mathbf{V}_j^{-1}(v) \bXjl  \bT_{\xi l } \big)
- \sum_{j=1}^n \brj' \mathbf{V}_j^{-1}(v) \bXjl \bT_{\xi l } \bXjl' \mathbf{V}_j^{-1}(v) \brj(v) \label{pdLsigma} \\
\noalign{and} 
\frac{\partial L}{\partial \rho_{\xi l}(k) } & =  \sigma_{\xi l}^2 \sum_{j=1}^n 
\mathrm{tr}\big( \bXjl' \mathbf{V}_j^{-1}(v) \bXjl  \bD_k  \big)
- \sigma_{\xi l}^2 \sum_{j=1}^n \brj' \mathbf{V}_j^{-1}(v) \bXjl \bD_k \bXjl' \mathbf{V}_j^{-1}(v) \brj (v) \label{pdLrho}
\end{align}
for $1\le k \le K-1$ and $ 1\le l \le L$. Based on \eqref{pdLsigma}-\eqref{pdLrho} and a suitable gradient-based optimization procedure, 
we obtained the ML estimators $\hat{\sigma}_{\xi l}^2 (v) $ and $\hat{\rho}_{\xi l}(k,v)$. 
We then define the aggregated correlation estimate $\hat{\rho}_{\xi l}(k)$
as the median of the $\hat{\rho}_{\xi l}(k,v)$ across voxels where the estimation was carried.

\section{Sampling distribution of the HRF estimators} 

Here we provide the theoretical justification of the large-sample approximation \eqref{eq: inference}  
to the sampling distribution of the estimators $\hat{\bbeta}(v)$ and $\hat{\bgamma}(v)$. 

First, the estimators used in this paper rely on standard statistical procedures whose consistency properties are well documented in the literature. 
 In step 1, the penalized least squares estimator $\hat{\mathbf{h}}(v)$ of the HRF coefficients $\gamma_{lk}(v)$ is consistent as $n\to\infty $ and $\lambda_0 \to 0$. 
Note that increasing the number of scans $T_j$ reduces the influence of the noise $\varepsilon$ on the estimation 
but not the sampling variability (subject effects $\xi$). In fact, the variance of the pilot estimator $\hat{\mathbf{h}}(v)$ is dominated by the sampling variability: it is of order 
 $\mathcal{O}(\sum_j T_j^2 / (\sum_j T_j)^2 )$, i.e.  $\mathcal{O}(1/n)$ if the $T_j$ are of comparable size. 
 Also, the parameter $\lambda_0$ governing the penalty on HRF shapes lying outside the null space $\boldsymbol{\Psi}$ must go to zero 
to render the pilot estimator asymptotically unbiased. 
 In step 2, the Yule-Walker estimators of the noise parameters $\sigma_{\varepsilon m}^2$ and $\boldsymbol{\theta}_{\varepsilon m}$ are consistent as $\max_j T_j \to \infty$. 
 The consistency of steps 3-5 derives from the large-sample properties of least squares- and maximum likeklihood estimators and from the consistency of the previous estimation steps. 
Note that the estimators of the noise and random effects parameters are averaged across subjects and space (except for the variance estimators $\hat{\sigma}_{\xi l}^2(v)$). As a consequence, their variance is generally small in comparison to the variance  of the HRF estimators.

We now turn to the large-sample distribution of $\hat{\bbeta}(v)$ and $\hat{\bgamma}(v)$. 
Assuming that for each subject, stimuli of each type are presented sufficiently often, 
the design matrices $\bXj$ are of order $\mathcal{O}(\sqrt{T_j})$ in norm 
and the matrix $\bM = \bM(v)$ is of order $\mathcal{O}(\sum_j T_j)$ in probability. 
Given the consistency of $\hat{\bgamma}(v)$, $\hat{\bV}_j(v)$, the normality of the data, 
and the fact that $\mathrm{Var}(\boldsymbol{\eta}) \approx \bM$, 
one can apply Slutsky's theorem and the Law of Large Numbers in \eqref{update beta gls} 
to obtain the large-sample approximation $ \hat{\bbeta}(v) \to N(\bbeta(v) ,  [  ( \bI_L \otimes  \bgamma(v) )' \bM    ( \bI_L \otimes  \bgamma(v)  )  ]^{-1} )$ 
(with $\hat{\bV}_j^{-1}(v)$ in $\bM$ replaced by $\hat{\bV}_j^{-1}(v)$) as $n,\sum_j T_j \to\infty$ and $\lambda \to 0$. 
Turning to the estimator $\hat{\bgamma}(v)$, simple algebraic manipulations in 
\eqref{update beta gls} and \eqref{update gamma gls} show that $\hat{C} = -n\lambda\, \hat{\bgamma}(v)' \bP \hat{\bgamma}(v)$. 
In other words, the terms $\hat{C}\bI_K$ and   $n\lambda \bP$ in \eqref{update gamma gls} 
are of order $\mathcal{O}(n \lambda)$ in probability and thus negligible in comparison to $\bM$. 
Applying the same arguments as with $\hat{\bbeta}(v)$ in \eqref{update beta gls}, 
we obtain the limit distribution  $ \hat{\bgamma}(v) \to N(\bgamma (v),    [  (  \bbeta (v) \otimes\bI_K   )'\bM  ( \bbeta (v) \otimes \bI_K ) ]^{-1}  )$.

\newpage

\begin{figure}
  \centering
    \includegraphics[width=0.9\textwidth]{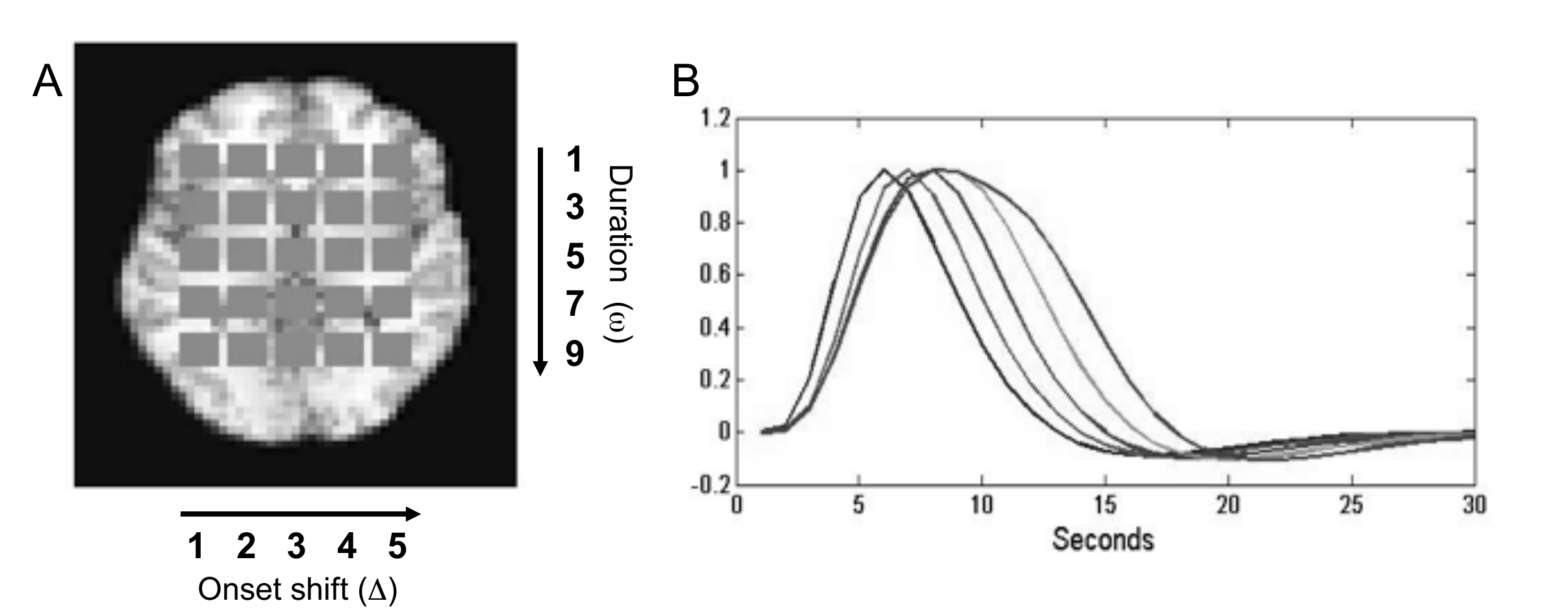}
\caption{{\small Overview of simulation set-up (A) A set of 25 equally sized squares were placed within a static brain image to represent regions of interest. BOLD signals were simulated based on different stimulus functions, which varied systematically across the squares in their onset and duration of neuronal activation. From left to right the onset of activation varied between the squares from the first to the fifth TR. From top to bottom, the duration of activation varied from one to nine TR in steps of two. (B)  The five HRFs with varying duration. The plot illustrates differences in time-to-peak and width attributable to changes in duration.}} \label{SimDesc}
\end{figure}

\begin{figure}
  \centering
    \includegraphics[width=1.0\textwidth]{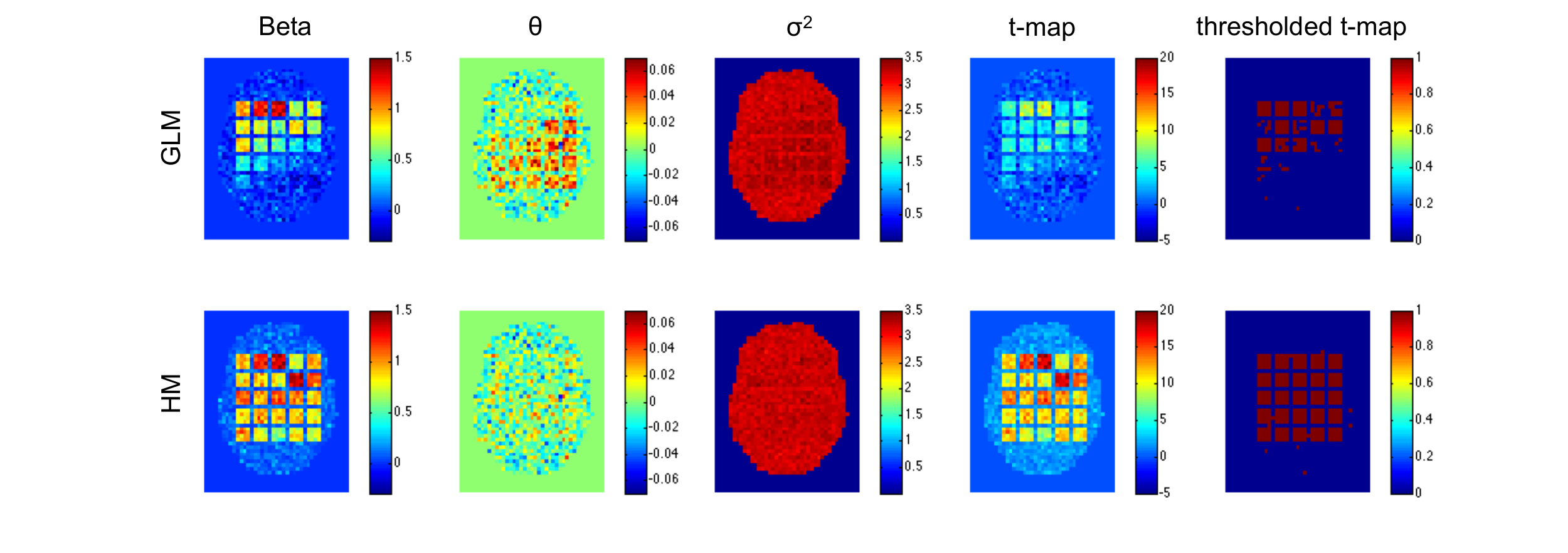}
\caption{{\small Results of the first simulation shown for the standard GLM/OLS approach (top row), and our hierarchical model (bottom row). From left-to-right the columns represent the estimated values of $\beta$, $\theta_{\varepsilon}$, $\sigma^2_{\varepsilon}$, $t$-map and thresholded $t$-map. }} \label{Sim1Res}
\end{figure}

\begin{figure}
  \centering
    \includegraphics[width=1.0\textwidth]{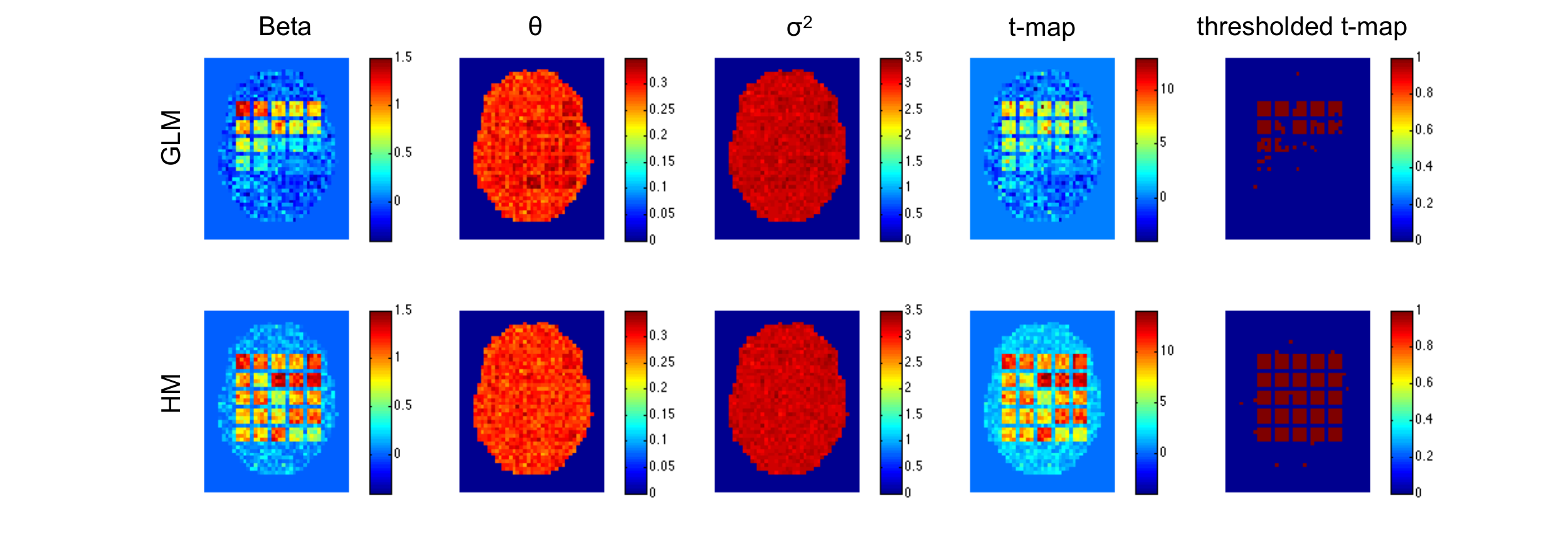}
\caption{{\small Results of the second simulation shown for the standard GLM/OLS approach (top row), and our hierarchical model (bottom row). From left-to-right the columns represent the estimated values of $\beta$, $\theta_{\varepsilon}$, $\sigma^2_{\varepsilon}$, the $t$-map and the thresholded $t$-map. }} \label{Sim2Res}
\end{figure}

\begin{figure}
  \centering
    \includegraphics[width=1.0\textwidth]{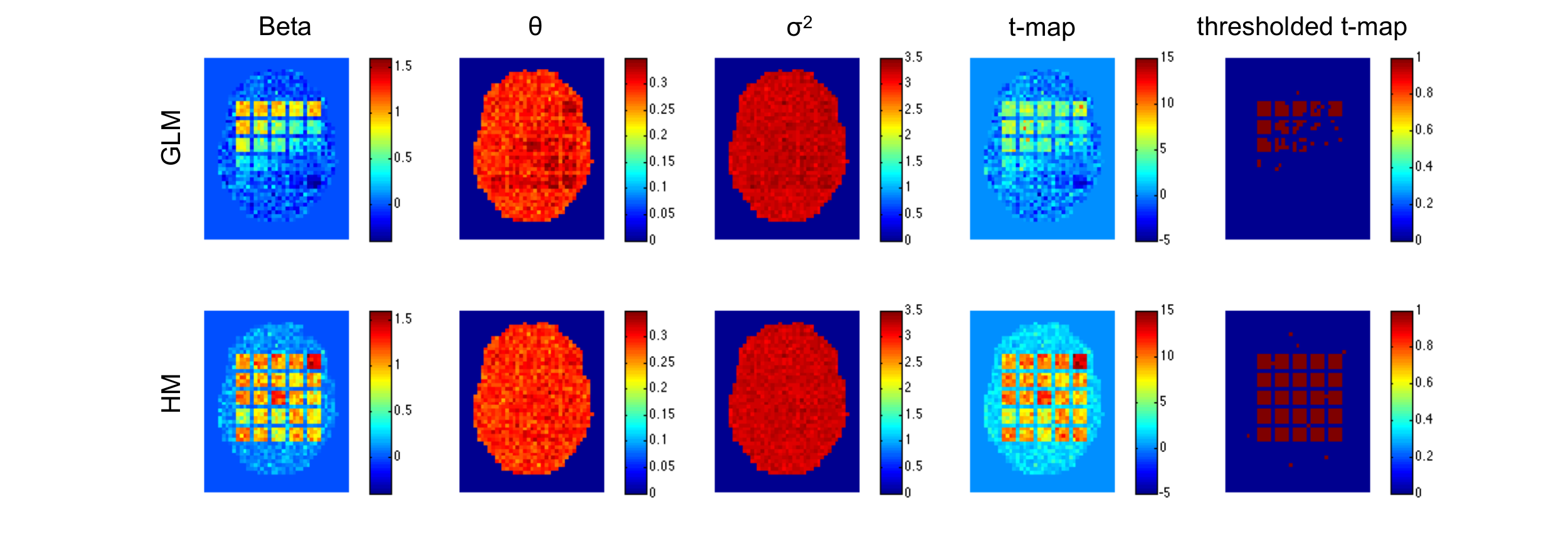}
\caption{{\small Results of the third simulation shown for the standard GLM/OLS approach (top row), and our hierarchical model (bottom row). From left-to-right the columns represent the estimated values of $\beta$,  $\theta_{\varepsilon}$, $\sigma^2_{\varepsilon}$, $t$-map and thresholded $t$-map. }} \label{Sim3Res}
\end{figure}

\begin{figure}
  \centering
    \includegraphics[width=0.7\textwidth]{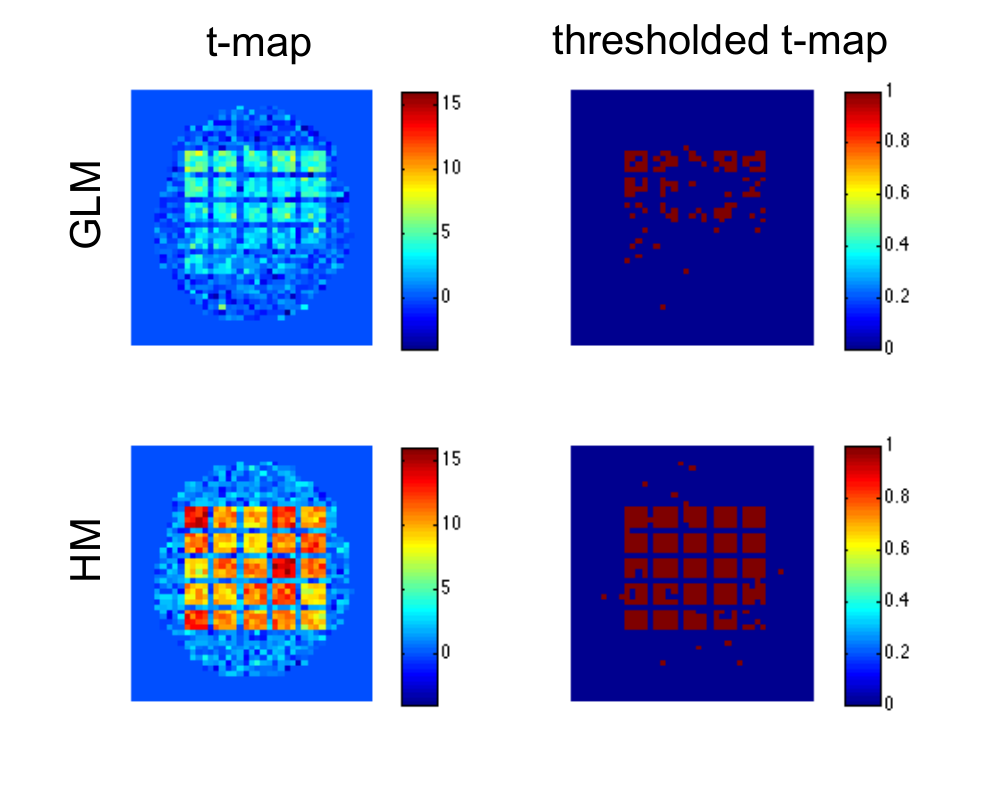}
\caption{{\small Results of the fourth simulation shown for the standard GLM/OLS approach (top row), and our hierarchical model (bottom row). From left-to-right the columns represent the estimated and thresholded $t$-map for testing whether the contrast between the two conditions was significantly different from 0.}} \label{Sim4}
\end{figure}

\begin{figure}
  \centering
    \includegraphics[width=0.7\textwidth]{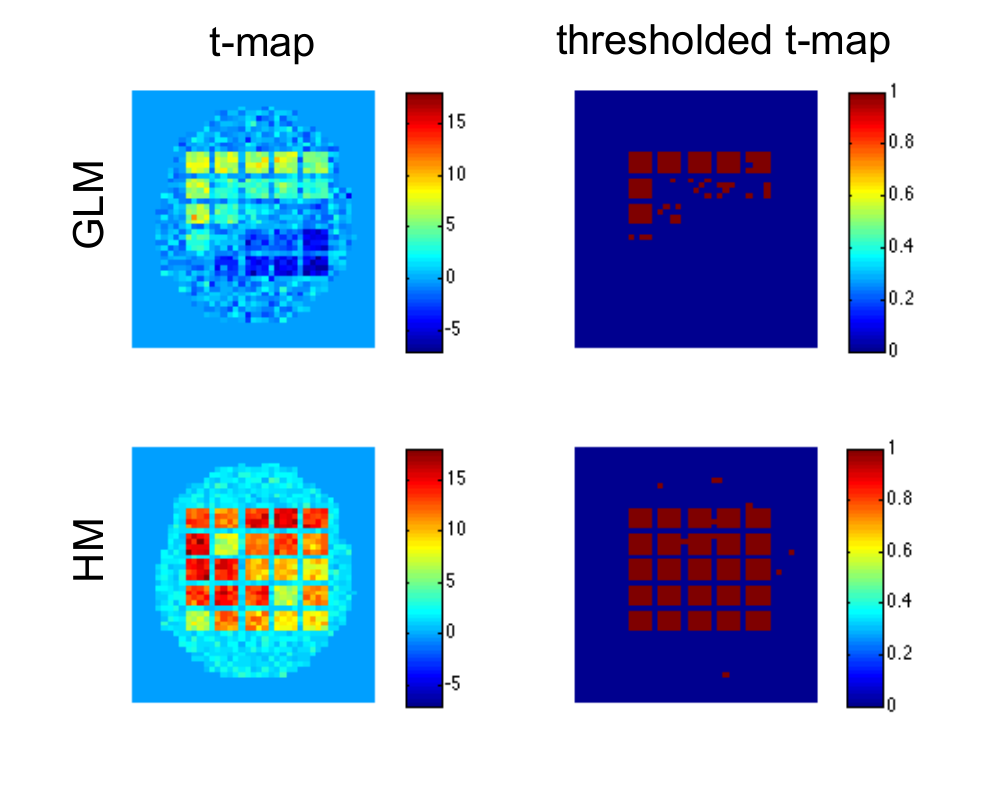}
\caption{{\small Results of the fifth simulation shown for the standard GLM/OLS approach (top row), and our hierarchical model (bottom row). From left-to-right the columns represent the estimated and thresholded $t$-map. }} \label{Sim5}
\end{figure}

\begin{figure}
  \centering
    \includegraphics[width=0.9\textwidth]{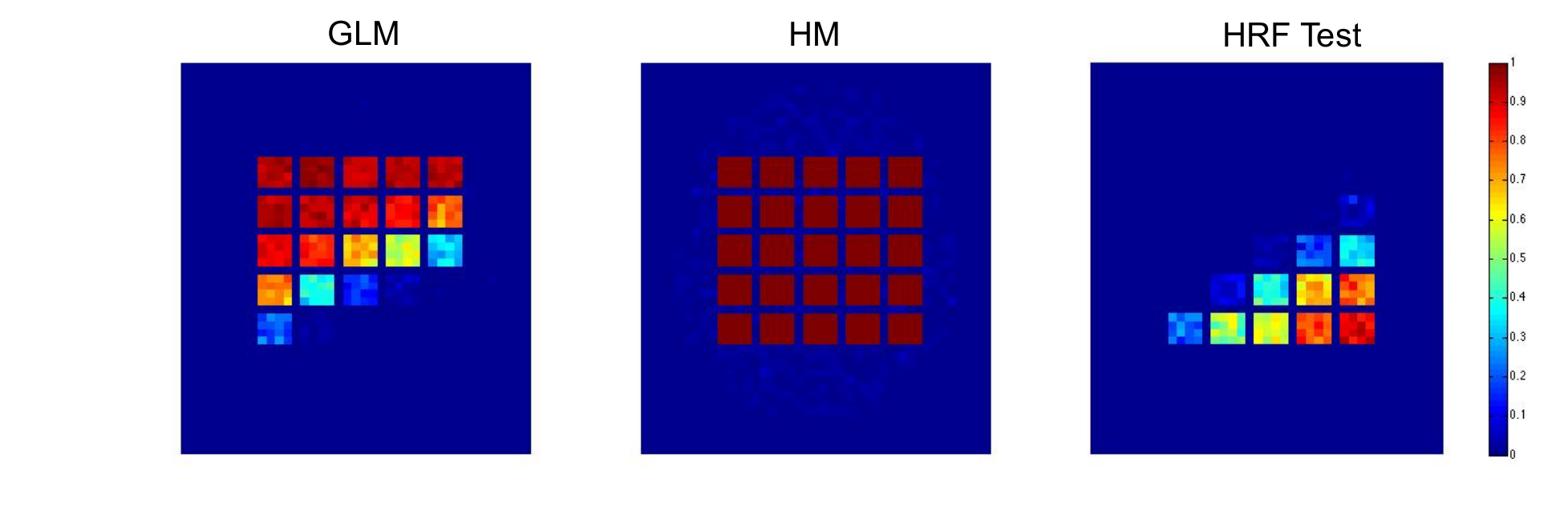}
\caption{{\small The results of $100$ replications of the first simulation. (A) The portion of times the standard GLM/OLS approach gave significant results in each voxel. (B) The same results for our hierarchical model. Clearly the hierarchical model is able to effectively separate signal from noise in a more consistent manner than the GLM/OLS. (C) The portion of times the estimated HRF deviated from the canonical form using our model.}} \label{Sim3b}
\end{figure}

\begin{figure}
  \centering
    \includegraphics[width=0.7\textwidth]{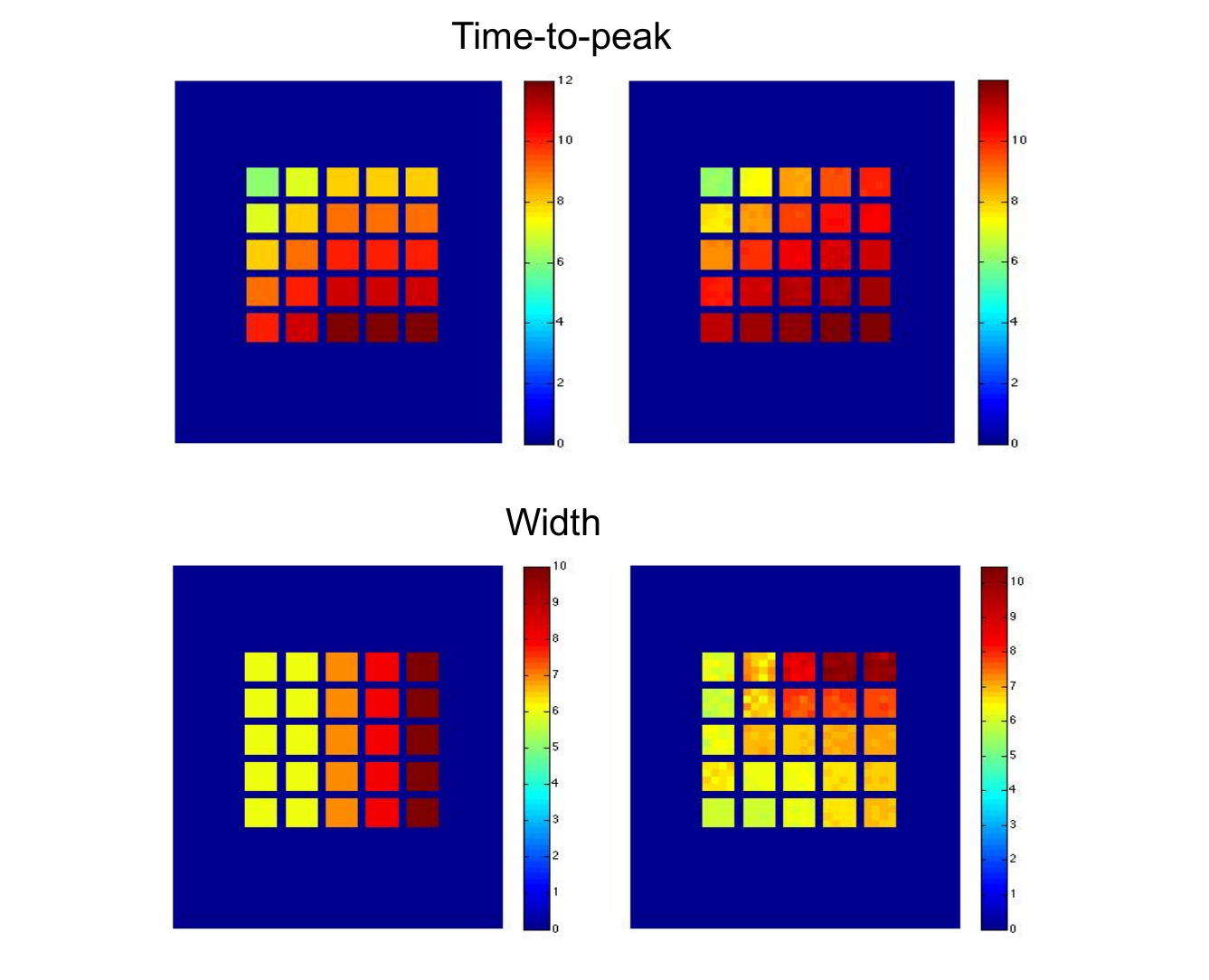}
\caption{{\small The results of $100$ replications of the first simulation.  (Top row)  The true and estimated values of the time-to-peak for the group-level HRF. (Bottom row) Same results for the width.}} \label{Sim3bTW}
\end{figure}

\begin{figure}[h]
  \centering
    \includegraphics[width=0.9\textwidth]{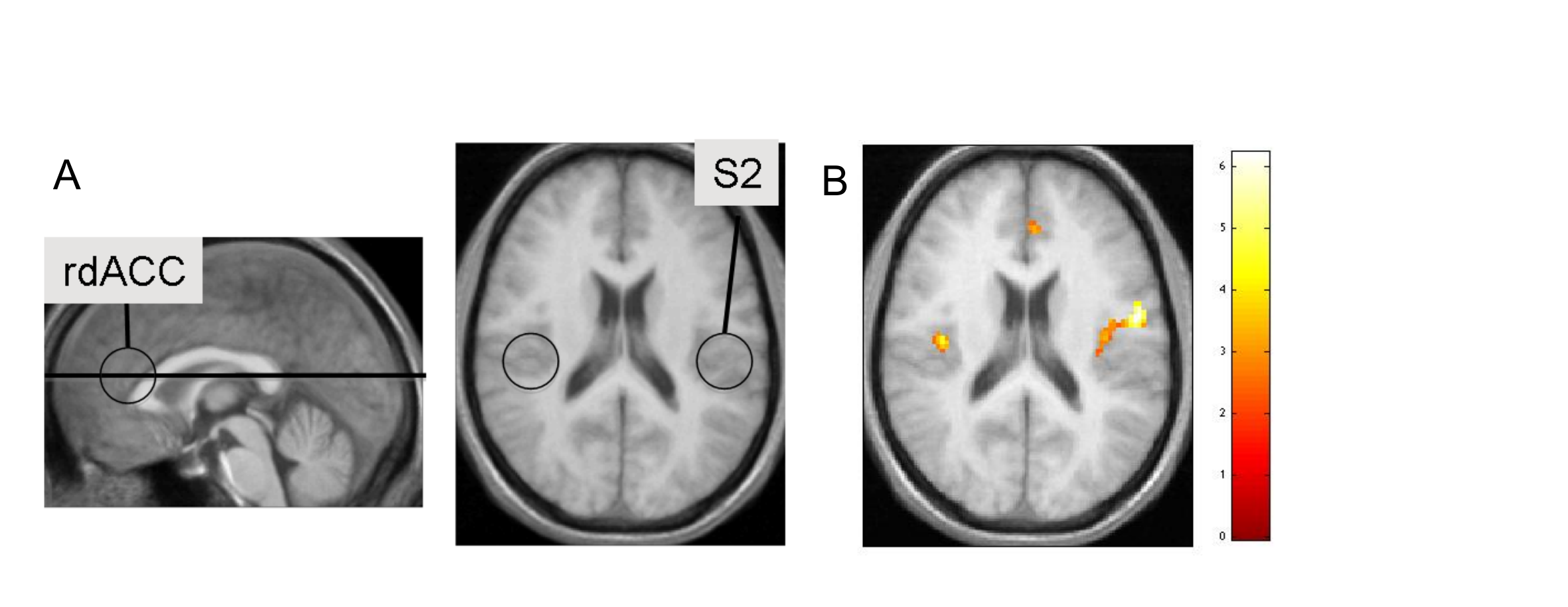}
\caption{{\small (A) The location of the slice and an illustration of areas of interest. Both rdACC and S2 are regions known to process pain intensity. 
(B) A statistical map obtained using the proposed herarchical model.  }} \label{Results}
\end{figure}

\begin{figure}[h]
  \centering
    \includegraphics[width=0.7\textwidth]{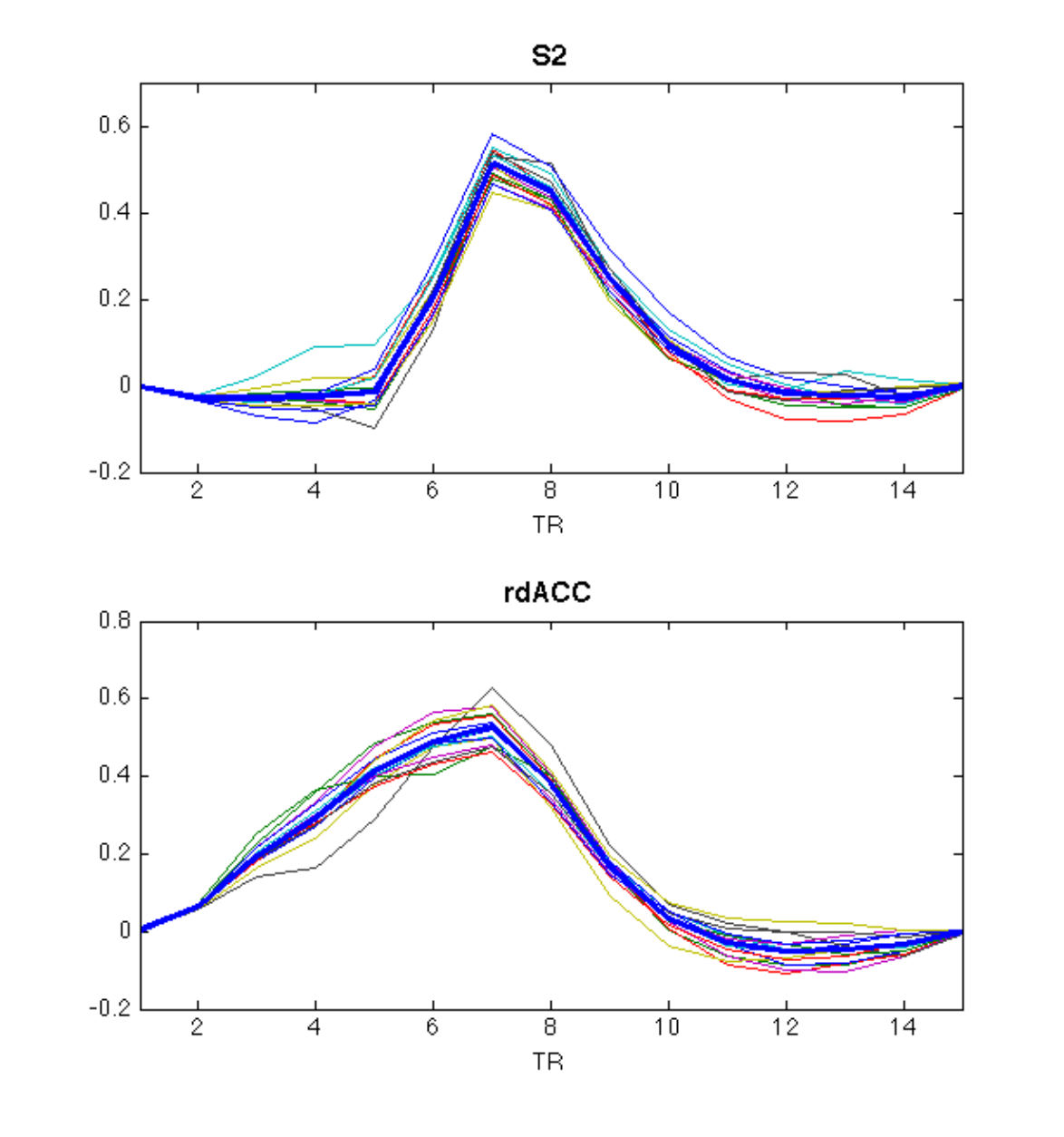}
\caption{{\small Estimates of the subject-specific HRF computed using voxels from the rdACC and S2. The group-level estimates are shown in bold.}} \label{ResultsHRF}
\end{figure}

\end{document}